\documentclass[12pt]{article}

\usepackage[titletoc,title]{appendix}
\usepackage[latin1]{inputenc}
\usepackage{amsmath}
\usepackage{amssymb}
\usepackage{graphicx}
\usepackage[margin=1.25in]{geometry}
\usepackage[bottom]{footmisc}
\usepackage{indentfirst}
\usepackage{endnotes}
\usepackage{mathabx}
\usepackage{enumerate}
\usepackage{graphicx}
\usepackage{rotating}
\usepackage{multirow}
\usepackage{booktabs,calc}
\usepackage{threeparttable}
\usepackage{float}
\usepackage{url}
\usepackage{epsfig}
\usepackage{caption,subcaption}
\usepackage{amsmath,amsfonts}
\usepackage{amsthm}
\usepackage{mathrsfs}
\usepackage{kbordermatrix}

\usepackage{enumitem}

\usepackage[onehalfspacing]{setspace}
\usepackage{apptools}

\usepackage{booktabs}
\usepackage{adjustbox}

\newcommand{\suchthat}{\;\ifnum\currentgrouptype=16 \middle\fi|\;}

\newtheorem{assm}{Assumption}
\newtheorem{theorem}{Theorem}
\newtheorem{prop}{Proposition}

\newtheorem{cor}{Corollary}

\newtheorem{example}{Example}
\newtheorem{defn}{Definition}

\theoremstyle{remark}

\usepackage{chngcntr}
\usepackage{graphicx}
\usepackage{amsmath}
\usepackage{amssymb}
\usepackage{parskip}
\usepackage{sgame}
\usepackage{color}
\usepackage{tikz}
\usetikzlibrary{trees,calc}

\usepackage{natbib}

\usepackage{dcolumn}                % aligning decimals
    \newcolumntype{d}[1]{D{.}{.}{#1}}

\begin{document}

\title{\large{Accurate Quality Elicitation in a Multi-Attribute Choice Setting}\thanks{
I am grateful for the guidance and encouragement of this project from John Rehbeck. I also thank Yaron Azrieli, Peter Caradonna, Paul J. Healy, Ian Krajbich, Jinkwon Lee, Kirby Nielsen, John Rehbeck, Jason Somerville, Charles Sprenger, Jason Tayawa, Ryan Webb, and participants at the OSU Theory/Experimental reading group, Sogang Behavioral/Experimental reading group, 17th EGSC at Washington University, 2022 North American ESA meeting, MEA 87th annual meeting, 2023 Midwest Economic Theory Conference, 2023 Econometric Society North America summer meeting, 2023 Caltech summer school and workshop, 2023 ESA World meeting, and 34th Stony Brook Game Theory Conference for helpful comments.
The experiment was determined exempt under IRB at Ohio State with the protocol number 2022E0907, and the Decision Science Collaborative at Ohio State supported the funding. All errors are mine.}}

\author{
    \small{Changkuk Im}  \\
    \small{Department of Economics} \\
    \small{The Ohio State University} \\
    \small{im.95@osu.edu}
}
\date{\small{ \today }}

\maketitle
\begin{abstract}
This paper studies how to accurately elicit quality for alternatives with multiple attributes. Two multiple price lists (MPLs) are considered: (i) m-MPL which asks subjects to compare an alternative to money, and (ii) p-MPL where subjects are endowed with money and asked whether they would like to buy an alternative or not. Theoretical results show that m-MPL requires fewer assumptions for accurate quality elicitation compared to p-MPL. Experimental evidence from a within-subject experiment using consumer products shows that switch points between the two MPLs are different, which suggests that quality measures are sensitive to the elicitation method.\\

\noindent\emph{JEL Classification Numbers:} C91, D11, D12 \\
\emph{Keywords:} Multi-attribute choice, Quality, Multiple price list, Consumer products, Initial endowment
\end{abstract}

\newpage
\setlength\parindent{24pt}

\section{Introduction}\label{sec:intro}
Quality is one of the attributes of interest in the marketing literature. Previous empirical studies measure quality in various ways, and measuring accurate quality values in these studies is essential to form correct conclusions. In this paper, quality means the subjective value of a product, and we study whether multiple price list (MPL) methods can accurately elicit quality.\footnote{MPLs are well-known experimental methods to elicit a subjective value of an alternative. They are broadly used in experimental studies investigating risk preferences \citep{holt2002risk}, time preferences \citep{andersen2006elicitation, andersen2008eliciting}, beliefs \citep{holt2016belief}, valuations for products \citep{anderson2007valuation}, and so on.}
This is not a trivial question in a multi-attribute choice setting since subjective quality is treated as a distinct attribute and multi-attribute choice models often allow menu-dependent context effects. Moreover, in an experimental setting, \cite{azrieli2018incentives, azrieli2020incentives} show that proper elicitation crucially depends on the theoretical assumptions that a researcher postulates. In this vein, the goal of this paper is to theoretically and experimentally examine how to accurately elicit quality using MPLs for multi-attribute choice models.

We focus on eliciting subjective quality in a multi-attribute choice setting where each alternative consists of multiple attributes such as quality, price, and/or monetary earnings in the spirit of \cite{bordalo2013salience}, \cite{kHoszegi2013model}, \cite{bushong2021model}, and \cite{landry2021pairwise}.\footnote{Multi-attribute choice models are widely explored in the marketing literature. Recent papers in economics investigate multi-attribute choice models under numerous domains including consumer, risky, intertemporal, and stochastic choices \citep{bordalo2012saliencein, bordalo2012salience, bordalo2013salience, bordalo2020memory, kHoszegi2013model, bushong2021model, landry2021pairwise, allen2022revealed}.} 
Specifically, we define a class of \emph{weighted separable attribute models} where the total utility of an alternative is a weighted sum of utility from attributes. This class of models nests several existing multi-attribute choice models including the range normalization models \citep{kHoszegi2013model, bushong2021model} and pairwise normalization models \citep{landry2021pairwise}. 

Two types of MPLs are considered in this paper.\footnote{In this paper, we interpret the Becker-DeGroot-Marschak (BDM) mechanism \citep{becker1964measuring} as an MPL since the decision problems in the BDM mechanism can be identically explained by the list format (see \cite{healy2018explaining}).} The first type is called \textit{m-MPL} where it asks a subject whether they prefer a product to a varying amount of money denoted by $m$ in Table~\ref{tab:mpl}. The second type is called \textit{p-MPL} where a subject first receives an initial endowment and then is asked whether they prefer to buy a product or not as price varies, where price is denoted by $p$ in Table~\ref{tab:mpl}. To elicit quality, one can find a \textit{switch point} which is defined by a monetary value in m-MPL or price in p-MPL where a subject switches from alternative $x$ to $y$ in Table~\ref{tab:mpl}. We say that an MPL \textit{accurately elicits} quality when the switch point equals subjective quality. 

\begin{table}
\caption{Decision problems in m-MPL and p-MPL}
\begin{center}
\begin{tabular}{lccc}
\toprule
MPL type & Alternative $x$                 & or & Alternative $y$ \\ \midrule
m-MPL     & Get Product X                &    & Get $\$m$    \\
p-MPL     & Buy Product X at price $\$p$ &    & Do nothing  \\
\bottomrule
\end{tabular}
\end{center}
\vspace{3mm}
\footnotesize{
%Example of m-MPL. 
%A switch point is defined by a question number where the subject switches their decision from Option A to B. 
\textit{Notes:} In m-MPL, if $x$ is chosen, then the subject obtains the product. The payment is not required. If $y$ is chosen, then the subject gets $\$m$. In p-MPL, if $x$ is chosen, then the subject buys the product by paying $\$p$ and keeps the remaining dollar amount, i.e., $\$(E-p)$ where $E$ denotes an initial endowment. If $y$ is chosen, then the subject keeps $\$E$.
}
\label{tab:mpl}
\end{table}

The goal of the theoretical part is to verify which set of assumptions guarantees accurate quality elicitation for each MPL. First, for any weighted separable attribute models, m-MPL accurately elicits quality under a mild assumption. To see why this is true, note that choosing the product in m-MPL gives marginal weighted utility from a quality attribute, and choosing the money in m-MPL gives marginal weighted utility from a money attribute. Assuming that quality and money attributes are expressed in the same unit, an injective marginal weighted utility function is necessary and sufficient to accurately elicit quality in m-MPL. 

On the other hand, p-MPL requires assumptions in addition to the injective assumption to guarantee accurate quality elicitation. The main reason is that when subjects receive an initial endowment in p-MPL, they may \textit{ignore} it or treat it as an attribute \textit{separate} from price. In these cases, prices are in a negative domain whereas quality values are in a positive domain. Thus, an assumption that connects gains and losses is required. It is also possible that subjects may \textit{combine} the initial endowment and price in a single attribute. In this case, an additional assumption that disentangles the initial endowment and price is required. 

Overall, theoretical results imply that m-MPL is able to accurately elicit quality for a \textit{wider} range of multi-attribute choice models compared to p-MPL. Moreover, assumptions that guarantee accurate quality elicitation for p-MPL depend on how subjects treat the initial endowment. Unless an experimenter can control this, further assumptions on how subjects treat the initial endowment are required to use p-MPL to accurately elicit quality. In contrast, m-MPL avoids this issue since an initial endowment is not present. 

It is useful to highlight how the theoretical findings predict quality elicitation using the two MPLs can differ. For example, given a range normalization model with a kinked utility function (e.g. gain-loss utility), p-MPL underestimates quality (see Example~\ref{ex:RN_kinked}). If the disparity is substantial, then closely investigating it is informative. If the disparity is negligible, then which MPLs used to measure subjective quality may be an unimportant issue despite the theoretical findings of this paper. For exploratory purposes, a within-subject experiment using consumer products is conducted to check whether the disparity exists and, if so, how large it is. 

The within-subject design allows us to examine the disparity at both aggregate and individual levels. The data show that, on average, subjective quality elicited by p-MPL is lower by 35\% relative to the quality elicited by m-MPL. While there exists heterogeneity across subjects, more than 70\% of subjects reported a lower quality value when p-MPL is used compared to when m-MPL is used. Further statistical analyses support that the disparity is substantial and robust in Section~\ref{sec:experiment}.

Both theoretical and experimental findings suggest that p-MPL may provide inaccurate quality elicitation in a multi-attribute choice setting. Yet, we should be cautious in concluding that p-MPL is obsolete. Indeed, p-MPL is essentially a measure of willingness to pay (WTP) for a product \citep{cunningham2013comparisons}. Thus, our findings suggest that quality and WTP may differ in general,\footnote{Whether these provide additional ability to distinguish between types of consumers and purchasing behavior is an open question. For instance, we may be able to identify consumers' preferences for loss aversion or risk aversion by looking at the disparity that is measured similarly to this paper or previous studies exploring willingness to accept and willingness to pay \citep{gachter2022individual, mrkva2020moderating, chapman2023willingness}.} and a researcher can choose an MPL depending on which information they want to use for their research question.

In Section~\ref{sec:discuss}, we relate the findings of this paper to those from the literature on the endowment effect, research from marketing, and research from consumer studies. First, inspired by \cite{kahneman1990experimental}, we connect the disparity observed in our experiment to the endowment effect. One of the experiments by \cite{kahneman1990experimental} decomposes the endowment effect into a reluctance to buy and a reluctance to sell. By interpreting that m-MPL is a measure of quality while p-MPL is a measure of WTP, our data show that the WTP is lower than the quality, which supports reluctance to buy. We also discuss how the wealth effect, framing effect, show-up fee, and other behavioral factors can be involved in the disparity.

Second, a marketing paper by \cite{park2008eliciting} proposes an upgrading method that measures marginal attribute values for a product with finitely many attributes. For instance, consider the decision problem of choosing a car with multiple attributes such as fuel economy, safety, comfort, and navigation system. Suppose that there are two completely identical vehicles except for one attribute. If there is a dollar amount that makes a decision maker indifferent between the superior vehicle and the inferior vehicle plus the money, then the upgrading method concludes that the dollar amount is equal to the marginal attribute value. We clarify the theoretical backgrounds of the upgrading method based on the findings of this paper.

Third, \cite{somerville2022range} proposes an experimental design to test multi-attribute choice models in a consumer context. The experiment consists of two steps: an elicitation step and a main choice task. In the elicitation step, subjective quality values of consumer products are elicited. The elicited values are used to proceed with precise comparative statics by varying the prices of the products in the main choice task. Since the comparative statics heavily depend on the elicited values, accurate quality elicitation in the elicitation step is essential. \cite{somerville2022range} uses the p-MPL type method in their elicitation step. As a comparison, we demonstrate the benefit of using m-MPL in this design.

There are many alternative methods that the marketing literature uses to estimate quality. For example, reports published by Consumer Union contain information such as opinions of products by consumers and experts, and they are used to measure quality when studying consumers' brand choice behavior \citep{hardie1993modeling}, firms' branding decisions \citep{montgomery1992risk}, and the relationship between price and quality \citep{gerstner1985higher}. \cite{parasuraman1988servqual} suggest a way to measure overall service quality by incorporating survey data asking about multiple dimensions of service. Average ratings of books from Amazon.com and Barnesandnoble.com are used as a proxy for the overall quality of books \citep{sun2012does, chevalier2006effect}. When studying consumers' behavior on a video-on-demand service, the quality of the digital signal is used as a proxy for the overall quality of the service \citep{nam2010effect, sriram2015service}. In this paper, we focus on accurate quality elicitation for multi-attribute choice models in an experimental setting.

As this paper compares different types of MPLs, it contributes to the methodological literature on MPLs \citep{andersen2006elicitation, sprenger2015endowment, collins2015response, brown2018separated, beauchamp2020measuring, hascher2021incentivized}. In terms of investigating MPLs in a multi-attribute choice setting, a close study is \cite{dertwinkel2022concentration} which shows how to elicit an attribute value while avoiding a behavioral bias in intertemporal choices under a range normalization model. In this paper, we study how to accurately elicit quality in a consumer context for a wider range of multi-attribute choice models. From a broader perspective, this paper is related to experimental methodology studies of \cite{azrieli2018incentives, azrieli2020incentives}. They verify assumptions that guarantee incentive compatible payment mechanisms when there are multiple tasks in an experiment. Similarly, this paper clarifies the required assumptions that guarantee accurate quality elicitation.

The remainder of the paper is organized as follows. Section~\ref{sec:framework} defines a class of multi-attribute choice models and clarifies the quality elicitation problem. Section~\ref{sec:result} provides the main theoretical results. Section~\ref{sec:experiment} presents the experimental design and evidence. Section~\ref{sec:discuss} discusses the findings by relating them to the existing literature. Section~\ref{sec:conclusion} concludes with final remarks.

\section{Framework}\label{sec:framework}
In this section, we define a class of multi-attribute choice models and the quality elicitation problem based on a multi-attribute choice setting in the spirit of \cite{bordalo2013salience}, \cite{kHoszegi2013model}, \cite{bushong2021model}, and \cite{landry2021pairwise}.

Let $x=(x_{1},\dots,x_{N})\in \mathbb{R}^{N}$ be an \textit{alternative} in which $x_{n} \in \mathbb{R}$ denotes an \textit{attribute value} on the $n$-th attribute. Attributes consisting of an alternative may depend on a choice context. For instance, if the context is buying a product, then the attributes would include the ``quality" and ``price" of the product.\footnote{In this paper, we assume that a decision maker encodes the overall quality of a product in a one-dimensional quality attribute. This is aligned with the literature studying multi-attribute choice models in a consumer context \citep{bordalo2013salience, landry2021pairwise, somerville2022range} and the marketing literature studying quality \citep{parasuraman1988servqual, sun2012does, nam2010effect}.} Note that all attribute values are expressed in the same unit such as monetary terms. Hence, if the common unit of measurement is in dollar value, then a quality attribute is interpreted as a decision maker's subjective value of a product in terms of dollars. 

Throughout the paper, we focus on a binary choice problem. Given two alternatives $x$ and $y$, a \textit{decision problem} or \textit{menu} is denoted by $\{x,y\}$. Given menu $\{x,y\}$, we assume that $x_{n}y_{n} \geq 0$ for all $n\in\{1,\dots,N\}$. This means that attribute values in the same attribute do not have different signs. For example, in a quality attribute, a quality value can be either positive or zero, but not negative. In a price attribute, a price value can be either negative or zero, but not positive.

Let $V:\mathbb{R}^{N}\times\mathbb{R}^{N}\rightarrow\mathbb{R}$ be an \textit{evaluation function} where its value indicates the evaluation of an alternative for a menu with two alternatives. The evaluation may depend on the other alternative in the menu, which allows menu-dependent context effects. Here, the first argument is the evaluated alternative and the second argument is the comparable alternative. For instance, $V(x,y)$ is the evaluation of $x$ in menu $\{x,y\}$. Throughout the paper, we write $V(x|\{x,y\})$ to emphasize that $x$ is the evaluated alternative in menu $\{x,y\}$. 

We focus on an evaluation function consisting of a utility function and a weight function. Let $u:\mathbb{R}\rightarrow\mathbb{R}$ be a \textit{utility function} that is weakly increasing, normalized to zero when an attribute value is zero so $u(0)=0$, and bounded so $|u(t)|<\infty$ for all $t \in \mathbb{R}$. %$\lim_{t\to\infty}u(t) < \infty$. 
Let  $w:\mathbb{R}\times\mathbb{R}\rightarrow\mathbb{R}_{+}$ be a nonnegative \textit{weight function} for an attribute that depends on the attribute values of all alternatives in the menu. We assume that it is bounded so $w(t,s)<\infty$ for all $t,s \in \mathbb{R}$. %$\lim_{t\to\infty}w(t,s) < \infty$, $\lim_{s\to\infty}w(t,s) < \infty$, and $\lim_{t,s\to\infty} w(t,s) < \infty$. 
Now, given a menu with two alternatives, we define a class of multi-attribute choice models in which the evaluation of an alternative is the sum of weighted utility indices.

\begin{defn}[Weighted separable attribute models]\label{def:weighted}
Let alternatives $x=(x_{1},\dots,x_{N})$ and $y=(y_{1},\dots,y_{N})$ in which $x_{n}y_{n}\geq0$ for all $n \in \{1,\dots,N\}$ be given. A model is called a weighted separable attribute model when its evaluation is the sum of weighted utility indices. Formally,
\begin{equation}\label{model:weighted}
\begin{split}
    V(x|\{x,y\}) &= \sum_{n=1}^{N} u(x_{n}) w(x_{n},y_{n}). 
    %\quad\text{and}
    %\\
    %V(y|\{y,x\}) &= \sum_{n=1}^{N} u(y_{n}) w(y_{n},x_{n}).
\end{split}\end{equation}
\end{defn}

The evaluation process of an alternative for a weighted separable attribute model can be understood as follows. The evaluation from a single attribute is the utility value times the weight given to that attribute from the context effects of attribute values of all alternatives. The final evaluation of an alternative sums the evaluations across attributes. Lastly, the decision maker chooses an alternative that has the highest final evaluation. Note that as attribute values are expressed in the same unit, the model has fixed utility and weight functions across attributes. Examples of these models are the range normalization models \citep{kHoszegi2013model, bushong2021model} and pairwise normalization models \citep{landry2021pairwise}.

To study the quality elicitation problem, it is useful to define a \textit{normalized weighted utility function} by $u^{0}:\mathbb{R}\rightarrow\mathbb{R}$ such that $u^{0}(t) = u(t)w(t,0)$ for all $t\in\mathbb{R}$. Its value indicates a weighted utility index when the context effect is normalized by zero comparable attribute value. Since $u(0)w(0,t)=0$, we can also think of it as $u^{0}(t) = u(t)w(t,0)-u(0)w(0,t)$. This can be interpreted as \textit{marginal weighted utility} that a decision maker attains from an attribute value of $t$ instead of 0 when both attribute values are available. In the MPL decision problems, the evaluation from the quality attribute can be summarized by a normalized weighted utility index, i.e., $u^{0}(q)$ where $q$ denotes the subjective quality of a product. Each MPL shows that $u^{0}(q)$ is equivalent to distinct evaluations from other attributes. Thus, for each MPL, the normalized weighted utility form helps us to understand the relationship between the quality and other attributes.

Now, we illustrate how alternatives in the MPLs can be represented in a multi-attribute choice setting. Here, we use the different MPLs to refer to both an \textit{elicitation method} and how attributes are \textit{encoded}. For example, the buying alternative in p-MPL can be differently represented as $(q,-p)$, $(q,-p,E)$, or $(q,E-p)$ with the relevant attributes. This is important because different results are derived depending on the type of MPLs and how attributes are encoded.

First, consider the m-MPL decision problem of choosing between $x$ and $y$ as in Table~\ref{tab:mpl} where $x$ is obtaining a product and $y$ is obtaining money. Assume that the first attribute is associated with quality and the second attribute is associated with money. Then the alternatives are written as $x=(q,0)$ and $y=(0,m)$ where $m$ denotes a dollar value that varies depending on a question number. Table~\ref{tab:menu} summarizes the representation of the alternatives from m-MPL.

\begin{table}
\caption{Representation of the alternatives from m-MPL and p-MPL in a multi-attribute choice setting}
\begin{center}
\begin{tabular}{lcccc}
\toprule
MPL type & m-MPL & p-MPL & p-MPL & p-MPL \\
\text{Endowment [Scenario]} &  No & \text{Yes [Ignore]} & \text{Yes [Separate]} & \text{Yes [Combine]} \\ 

\midrule
$x$ & $(q,0)$ & $(q,-p)$ & $(q,-p,E)$ & $(q,E-p)$ \\
$y$ & $(0,m)$ & $(0,0)$ & $(0,0,E)$ & $(0,E)$ \\ 
\midrule
1st attribute & Quality & Quality & Quality & Quality \\
2nd attribute & Money & Price & Price & Earnings \\
3rd attribute & - & - & Endowment & - \\
\bottomrule
\end{tabular}
\end{center}
\vspace{3mm}
\footnotesize{\textit{Notes:} Endowment denotes whether it requires an initial endowment. Scenario denotes how subjects treat the initial endowment during the decision-making process. Ignore scenario is when a subject ignores the initial endowment. Separate scenario is when a subject considers the initial endowment but separates it from price. Combine scenario is when a subject combines the initial endowment and price in a single attribute.}
\label{tab:menu}
\end{table}

To elicit quality with m-MPL, one can find a \textit{switch point}. To understand a switch point, suppose that monetary values in m-MPL are listed in ascending order. Then a subject is expected to choose $x$ for some initial questions, switch to $y$ at some question number, and choose $y$ for the remaining questions. A switch point is defined by a monetary value where a subject switches their decision from $x$ to $y$.

We say that m-MPL \textit{accurately elicits} quality given a model when the subjective quality equals the monetary value at the switch point assuming that a given model is true. Our goal is to verify which set of assumptions guarantees accurate quality elicitation given a class of weighted separable attribute models.

Suppose that a subject switches a choice from $x$ to $y$ at the dollar value of $m^{*}$. Then m-MPL tells us that $x$ and $y$ are indifferent at the switch point. However, this does not necessarily mean that the quality is equal to the dollar value at the switch point, i.e., $q=m^{*}$. From the representation of the alternatives in m-MPL, we find that the context effect for each attribute can be normalized due to zero comparable attribute value. Thus, for any weighted separable attribute model, the evaluation of each alternative in m-MPL can be simplified into a normalized weighted utility form. More specifically, we have $u^{0}(q) = V(x|\{x,y\})=V(y|\{y,x\}) = u^{0}(m^{*})$ at the switch point. To conclude that $q=m^{*}$, the normalized weighted utility function must be injective. Motivated by this, we introduce the first assumption.

\begin{enumerate}[label=A\theenumi,font=\itshape]
    \item (Injective) If $t>s>0$, then $u^{0}(t) \neq u^{0}(s)$.\label{assm:injective}
\end{enumerate}

The injective assumption of \ref{assm:injective} implies that each positive attribute value has a unique normalized weighted utility index. Here, we emphasize that the injective assumption is associated only with the positive domain. Thus, any strictly increasing normalized weighted utility function on the positive domain would satisfy the assumption. Proposition~\ref{prop:mMPL}  shows that the injective assumption alone is necessary and sufficient to guarantee accurate quality elicitation in m-MPL.

Next, consider the p-MPL decision problem of choosing between $x$ and $y$ as in Table~\ref{tab:mpl} where $x$ is buying a product and $y$ is doing nothing. Let $E$ be an initial endowment that a subject receives before making a decision. We can think of three different scenarios based on how subjects treat the initial endowment during the decision-making process. 

The first scenario is when subjects \textit{ignore} the initial endowment. Assume that the first attribute is associated with quality and the second attribute is associated with price. Then the alternatives are written as $x=(q,-p)$ and $y=(0,0)$ where $p$ denotes the price of the product that varies depending on a question number. The second scenario is when subjects consider the initial endowment but \textit{separate} it from price. Here, we assume that the first attribute is associated with quality, the second attribute is associated with price, and the third attribute is associated with endowment. In this case, the alternatives are written as $x=(q,-p,E)$ and $y=(0,0,E)$. The third scenario is when subjects \textit{combine} the initial endowment and price. Here, we assume that the first attribute is associated with quality and the second attribute is associated with earnings. In this case, the alternatives are written as $x=(q,E-p)$ and $y=(0,E)$. Table~\ref{tab:menu} summarizes the representations of the alternatives from p-MPL.

Similarly, to elicit quality with p-MPL, one can find a \textit{switch point} defined by a price where a subject switches their decision from $x$ to $y$ when prices are listed in ascending order. We say that p-MPL \textit{accurately elicits} quality given a model when the subjective quality is equal to the price at the switch point assuming that a given model is true. 

Suppose that a subject switches a choice from $x$ to $y$ at the price of $p^{*}$. In all scenarios, p-MPL tells us that $x$ and $y$ are indifferent at the switch point. However, this does not necessarily mean that the quality is equal to the price at the switch point, i.e., $q=p^{*}$. Unlike the m-MPL case, accurately eliciting quality with p-MPL is more complex because multiple scenarios are involved and the context effects may hinder accurate quality elicitation. Example~\ref{ex:RN_kinked} shows a case in which p-MPL cannot accurately elicit quality in the first and the second scenarios.

\begin{example}\label{ex:RN_kinked}
Consider a range normalization model that evaluates alternative $x$ in menu $\{x,y\}$ by $V_{RN}(x|\{x,y\}) = \sum_{n=1}^{N} u(x_{n})w_{RN}(x_{n},y_{n})$ where
\begin{equation*}\label{model:RD}
    w_{RN}(t,s) = 
    \begin{cases}
    |u(t) - u(s)|^{\gamma} &\text{ for } u(t) \neq u(s) \\
    0 &\text{ for } u(t) = u(s)
\end{cases}\end{equation*}
with $\gamma>-1$. Consider a kinked utility function such as
\begin{equation*}u(t)=\begin{cases}
    t \quad &\text{for } t\geq0\\
    \lambda t \quad &\text{for } t<0
    \end{cases}\end{equation*}
where $\lambda>1$. Here, $\lambda>1$ captures loss aversion. A range normalization model with a kinked utility function is nested in a class of weighted separable attribute models. Note that the associated normalized weighted utility function is $u^{0}(t) = t^{1+\gamma}$ for $t\geq0$ and $u^{0}(t)=-(-\lambda t)^{1+\gamma}$ for $t<0$.

Consider the first scenario when subjects ignore the initial endowment. In this case, the switch point from p-MPL is always smaller than the quality since $\lambda>1$. Formally, $u^{0}(q) + u^{0}(-p^{*}) = V_{RN}((q,-p^{*})|\cdot) = V_{RN}((0,0)|\cdot) = 0$ where the dots refer to corresponding menus implies $q = \lambda p^{*} > p^{*}$ where $p^{*}$ is the switch point and $\lambda p^{*}$ is the model-implied quality. We obtain the same result for the second scenario when subjects separate the initial endowment from price, i.e., $V_{RN}((q,-p^{*},E)|\cdot) = V_{RN}((0,0,E)|\cdot)$ in which the dots refer to corresponding menus.
\end{example}

The reason p-MPL cannot accurately elicit quality in Example~\ref{ex:RN_kinked} is that the normalized weighted utility function is not symmetric on gains and losses. Motivated by Example~\ref{ex:RN_kinked}, we introduce the second assumption.

\begin{enumerate}[label=A\theenumi,font=\itshape]
\setcounter{enumi}{1}
    \item (Symmetry) For any $t>0$, we have $u^{0}(t) = -u^{0}(-t)$.\label{assm:sym}
\end{enumerate}

The symmetry assumption of \ref{assm:sym} implies that given two attribute values with different signs but the same absolute value, their magnitudes of the normalized weighted utility index are the same. In other words, it guarantees that gain and loss domains are symmetric. Naturally, the symmetry assumption is associated with both negative and positive domains. If $\lambda=1$ in Example~\ref{ex:RN_kinked}, then the utility function becomes linear and the symmetry assumption is satisfied. Proposition~\ref{prop:pMPL12} shows that the injective and the symmetry assumptions are necessary and sufficient to guarantee accurate quality elicitation with p-MPL in the first and the second scenarios.

The following example shows a case in which p-MPL cannot accurately elicit quality in the third scenario. This example motivates an additional linearity assumption on the normalized weighted utility function. 

\begin{example}\label{ex:RN_power}
Consider the range normalization model as defined in Example~\ref{ex:RN_kinked}. Now, instead of a kinked utility function, consider a power utility function such as 
\begin{equation*}u(t)=\begin{cases}
    t^{\alpha} \quad&\text{for } t\geq0\\
    -|t|^{\alpha} \quad&\text{for } t<0
\end{cases}\end{equation*}
where $\alpha>0$ and $\alpha \neq 1$. A range normalization model with a power utility function is also nested in a class of weighted separable attribute models. Note that the associated normalized weighted utility function is $u^{0}(t)=t^{\alpha(1+\gamma)}$ for $t\geq0$ and $u^{0}(t) = -(-t)^{\alpha(1+\gamma)}$ for $t<0$.

Consider the third scenario when subjects combine the initial endowment and price. At the switch point, we have $q^{\alpha}q^{\alpha \gamma} + (E-p^{*})^{\alpha}(E^{\alpha}-(E-p^{*})^{\alpha})^{\gamma} = V_{RN}((q,E-p^{*})|\cdot) = V_{RN}((0,E)|\cdot) = E^{\alpha} (E^{\alpha}-(E-p^{*})^{\alpha})^{\gamma}$ in which the dots refer to corresponding menus. This implies that $q = (E^{\alpha}-(E-p^{*})^{\alpha})^{\frac{1}{\alpha}}$. If $\alpha \in (0,1)$, then the switch point is greater than the model-implied quality, i.e., $q = (E^{\alpha}-(E-p^{*})^{\alpha})^{\frac{1}{\alpha}}<p^{*}$. If $\alpha>1$, then the switch point is smaller than the model-implied quality, i.e., $q = (E^{\alpha}-(E-p^{*})^{\alpha})^{\frac{1}{\alpha}}>p^{*}$.
\end{example}

The reason p-MPL cannot accurately elicit quality in Example~\ref{ex:RN_power} is that the curvature from the power utility function prevents $u(E-p^{*})w(E-p^{*},E)$ and $u(E)w(E,E-p^{*})$ terms from being combined into a normalized weighted utility form. Motivated by Example~\ref{ex:RN_power}, we introduce the third assumption.

\begin{enumerate}[label=A\theenumi,font=\itshape]
\setcounter{enumi}{2}
    \item (Linearity) For any $t>s>0$,  we have $u^{0}(t-s) = u(t)w(t,s) - u(s)w(s,t)$.\label{assm:linear}
\end{enumerate}

The linearity assumption of \ref{assm:linear} can be understood as follows. Recall that we can interpret the normalized weighted utility as marginal weighted utility. Then the left-hand side of the equation is the marginal weighted utility that a decision maker attains from an attribute value of $t-s$ instead of 0 when both attribute values are available. Similarly, the right-hand side is the marginal weighted utility that a decision maker attains from $t$ instead of $s$ when both attribute values are available. Therefore, the assumption implies constant marginal weighted utility whenever the distance between two attribute values is the same. Note that the linearity assumption is associated only with the positive domain. If $\alpha=1$ in Example~\ref{ex:RN_power}, then the utility function becomes linear and the linearity assumption is satisfied. Proposition~\ref{prop:pMPL3} shows that the injective and the linearity assumptions are necessary and sufficient to guarantee accurate quality elicitation with p-MPL in the third scenario.

Note that while the symmetry assumption is not satisfied in Example~\ref{ex:RN_kinked}, the model in the example satisfies the linearity assumption. While the linearity assumption is not satisfied in Example~\ref{ex:RN_power}, the model in the example satisfies the symmetry assumption. The range normalization model with a linear utility function (e.g., $\lambda=1$ in Example~\ref{ex:RN_kinked} or $\alpha=1$ in Example~\ref{ex:RN_power}) satisfies both the symmetry and the linearity assumptions. These together show that the symmetry and the linearity assumptions are neither mutually exclusive nor imply the other. This is because the symmetry assumption is associated with both positive and negative domains, whereas the linearity assumption is associated only with the positive domain.

\section{Theoretical Results}\label{sec:result}
The goal of this section is to verify a set of assumptions on a class of weighted separable attribute models of (\ref{model:weighted}) that guarantees accurate quality elicitation for each MPL. Results for a more general class of multi-attribute choice models that nests the contextual concavity models of \cite{kivetz2004alternative} are presented in Appendix~\ref{appendix:thm}. 

Recall from the examples in the previous section that accurate quality elicitation is affected by which MPL is used and how subjects treat the initial endowment when it is given. We first prove Proposition~\ref{prop:mMPL} which shows that given two-attribute alternatives each consisting of one positive attribute value and one zero attribute value, a weighted separable attribute model connects distinct attributes one to one under the injective assumption.

\begin{prop}\label{prop:mMPL}
Let two-attribute alternatives $x=(x_{1},0)$ and $y=(0,y_{2})$ with $x_{1},y_{2}>0$ be given. Suppose that a decision maker chooses according to a weighted separable attribute model of (\ref{model:weighted}). Then the following two statements are equivalent.
\begin{enumerate}[label=(\alph*)]
    \item The model satisfies \ref{assm:injective} (Injective).
    \item $V(x|\{x,y\}) = V(y|\{y,x\})$ if and only if $x_{1}=y_{2}$.
\end{enumerate}
The result holds even when the alternatives have additional attributes with constants, for example, $x=(x_{1},0,c_{3},\dots,c_{N})$ and $y=(0,y_{2},c_{3},\dots,c_{N})$ in which $c_{3},\dots,c_{N} \in \mathbb{R}$.
\end{prop}

\begin{proof}[Proof of Proposition~\ref{prop:mMPL}] [(a) $\implies$ (b)] First, by definition, $V(x|\{x,y\})=V(y|\{y,x\})$ is equivalent to $u(x_{1})w(x_{1},0) = u(y_{2})w(y_{2},0)$. This can be rewritten as $u^{0}(x_{1}) = u^{0}(y_{2})$. Due to the injective assumption, we conclude that $x_{1}=y_{2}$. Next, suppose that $x_{1}=y_{2}$. Then we have $u^{0}(x_{1}) = u^{0}(y_{2})$, which implies that $V(x|\{x,y\}) = V(y|\{y,x\})$.

[(b) $\implies$ (a)] Let $t>s>0$ be given. Consider $x=(t,0)$ and $y=(0,s)$. By (b), we have $u^{0}(t) = u(t)w(t,0) = V(x|\{x,y\}) \neq V(y|\{y,x\}) = u(s)w(s,0) = u^{0}(s)$.

The same logic can be applied to prove the case when the alternatives have additional attributes with constants as the model is additively separable.
\end{proof}

The proof of the sufficiency part is straightforward as discussed in Section~\ref{sec:framework}. Due to zero attribute values, the context effect for each attribute can be normalized for any weighted separable attribute model. Thus, the evaluation of each alternative is rewritten by a normalized weighted utility function. Then the injective assumption connects the two attribute values from distinct attributes. 

As a set of alternatives in the proposition generalizes the m-MPL decision problem, the result indicates $q=m^{*}$ in the quality elicitation problem. Therefore, Proposition~\ref{prop:mMPL} implies that the injective assumption alone is necessary and sufficient for m-MPL to accurately elicit quality for any weighted separable attribute model. Corollary~\ref{cor:mMPL} records the sufficiency part.

\begin{cor}\label{cor:mMPL}
Suppose that a subject chooses according to a weighted separable attribute model of (\ref{model:weighted}) satisfying \ref{assm:injective} (Injective). If the subject considers quality and money attributes, then m-MPL accurately elicits quality. Formally,
    \begin{align*}
        V((q,0)|\{(q,0),(0,m^{*})\}) &= V((0,m^{*})|\{(0,m^{*}),(q,0)\})\\
        \text{ if and only if } q&=m^{*}.
    \end{align*}
\end{cor}

Proposition~\ref{prop:mMPL} also states that the result holds even when there are additional attributes with constants since the model is additively separable. This general result is useful, for example, to understand how an upgrading method proposed by \cite{park2008eliciting} works when an alternative has finitely many attributes (see Section~\ref{sec:discuss_update}). For similar reasons, the following propositions also state the results when there are additional attributes with constants.

Now, we examine the quality elicitation problem for the p-MPL cases. Recall that we can think of three different scenarios of how subjects treat the initial endowment: (i) when subjects ignore it, (ii) when subjects consider it but separate it from price, and (iii) when subjects combine it with price. To study the first and the second scenarios, we prove Proposition~\ref{prop:pMPL12} which shows that given two alternatives where one consists of two attribute values with different signs and the other one consists of zero attribute values, a weighted separable attribute model connects distinct attributes one to one under the injective and the symmetry assumptions.

\begin{prop}\label{prop:pMPL12}
Let two-attribute alternatives $x=(x_{1},x_{2})$ and $y=(0,0)$ with $x_{1}x_{2}<0$ be given. Suppose that a decision maker chooses according to a weighted separable attribute model of (\ref{model:weighted}). Then the following two statements are equivalent.
\begin{enumerate}[label=(\alph*)]
    \item The model satisfies \ref{assm:injective} (Injective) and \ref{assm:sym} (Symmetry).
    \item $V(x|\{x,y\}) = V(y|\{y,x\})$ if and only if $x_{1}=-x_{2}$.
\end{enumerate}
The result holds even when the alternatives have additional attributes with constants, for example, $x=(x_{1},x_{2},c_{3},\dots,c_{N})$ and $y=(0,0,c_{3},\dots,c_{N})$ in which $c_{3},\dots,c_{N} \in \mathbb{R}$.
\end{prop}

\begin{proof}[Proof of Proposition~\ref{prop:pMPL12}] 
[(a) $\implies$ (b)] First, by definition, $V(x|\{x,y\})=V(y|\{y,x\})$ is equivalent to $u(x_{1})w(x_{1},0) + u(x_{2})w(x_{2},0) = 0$. After rearranging it, this can be rewritten as $u^{0}(x_{1}) = u^{0}(-x_{2})$ due to the symmetry assumption. By the injective assumption, we conclude that $x_{1}=-x_{2}$. Next, suppose that $x_{1}=-x_{2}$. Then we have $u^{0}(x_{1}) = u^{0}(-x_{2})$, which implies that $V(x|\{x,y\})=V(y|\{y,x\})$. The same logic can be applied to prove the case when the alternatives have additional attributes with constants as the model is additively separable.

[(b) $\implies$ (a)] See Theorem~\ref{thm:pMPL12} in Appendix~\ref{appendix:thm}.
\end{proof}

The proof of the sufficiency part shows that the evaluation of the alternative consisting of positive and negative attribute values is the sum of their normalized weighted utility indices since the comparable alternative has zero attribute values. Then the injective and the symmetry assumptions together connect the positive and the negative attribute values without any distortion. The result holds even when the alternatives have additional attributes with constants due to the additive separability.

A set of two-attribute alternatives in Proposition~\ref{prop:pMPL12} generalizes the p-MPL decision problem in the first scenario. A set of three-attribute alternatives with a constant in an additional attribute from Proposition~\ref{prop:pMPL12} generalizes the p-MPL decision problem in the second scenario. Hence, Proposition~\ref{prop:pMPL12} implies $q=p^{*}$ in the quality elicitation problem with p-MPL under those scenarios. In other words, the injective and the symmetry assumptions together guarantee that p-MPL accurately elicits quality when subjects ignore the initial endowment or consider it but separate from price, which we record in Corollary~\ref{cor:pMPL12}.

\begin{cor}\label{cor:pMPL12}
Suppose that a subject chooses according to a weighted separable attribute model of (\ref{model:weighted}) satisfying \ref{assm:injective} (Injective) and \ref{assm:sym} (Symmetry). 
\begin{enumerate}
    \item If the subject ignores the initial endowment so considers quality and price attributes, then p-MPL accurately elicits quality. Formally,
    \begin{align*}
        V((q,-p^{*})|\{(q,-p^{*}),(0,0)\}) &= V((0,0)|\{(0,0),(q,-p^{*})\})\\
        \text{ if and only if } q&=p^{*}.
    \end{align*}
    \item If the subject treats the initial endowment separately from price and quality so considers quality, price, and endowment attributes, then p-MPL accurately elicits quality. Formally,
    \begin{align*}
        V((q,-p^{*},E)|\{(q,-p^{*},E),(0,0,E)\}) &= V((0,0,E)|\{(0,0,E),(q,-p^{*},E)\})\\
        \text{ if and only if } q&=p^{*}.
    \end{align*}
\end{enumerate}
\end{cor}

Finally, we examine the last scenario case when subjects combine the initial endowment and price in a single attribute. To this end, we prove Proposition~\ref{prop:pMPL3} which shows that given two-attribute alternatives where all of the attribute values are positive except one zero attribute value, a weighted separable attribute model connects the sum of attribute values of each alternative equivalently under the injective and the linearity assumptions.

\begin{prop}\label{prop:pMPL3}
Let two-attribute alternatives $x=(x_{1},x_{2})$ and $y=(0,y_{2})$ with $x_{1},x_{2},y_{2}>0$ and $y_{2}>x_{2}$ be given. Suppose that a decision maker chooses according to a weighted separable attribute model of (\ref{model:weighted}). Then the following two statements are equivalent.
\begin{enumerate}[label=(\alph*)]
    \item The model satisfies \ref{assm:injective} (Injective) and \ref{assm:linear} (Linearity).
    \item $V(x|\{x,y\}) = V(y|\{y,x\})$ if and only if $x_{1}+x_{2}=y_{2}$.
\end{enumerate}
The result holds even when the alternatives have additional attributes with constants, for example, $x=(x_{1},x_{2},c_{3},\dots,c_{N})$ and $y=(0,y_{2},c_{3},\dots,c_{N})$ in which $c_{3},\dots,c_{N} \in \mathbb{R}$.
\end{prop}

\begin{proof}[Proof of Proposition~\ref{prop:pMPL3}]
[(a) $\implies$ (b)] First, by definition, $V(x|\{x,y\})=V(y|\{y,x\})$ is equivalent to $u(x_{1})w(x_{1},0) + u(x_{2})w(x_{2},y_{2}) = u(y_{2})w(y_{2},x_{2})$. After rearranging it, this can be rewritten as $u^{0}(x_{1}) = u^{0}(y_{2}-x_{2})$ due to the linearity assumption. By the injective assumption, we conclude that $x_{1} + x_{2} = y_{2}$. Next, suppose that $x_{1} + x_{2} = y_{2}$. Then we have $u^{0}(x_{1}) = u^{0}(y_{2}-x_{2})$, which implies that $V(x|\{x,y\})=V(y|\{y,x\})$. The same logic can be applied to prove the case when the alternatives have additional attributes with constants as the model is additively separable.

[(b) $\implies$ (a)] See Theorem~\ref{thm:pMPL3} in Appendix~\ref{appendix:thm}.
\end{proof}

The proof of the sufficiency part shows that while the context effect is involved in the second attribute, we can combine $x_{2}$ and $y_{2}$ in a normalized weighted utility form due to the linearity assumption. Then the injective assumption connects the sums of attribute values, i.e., $x_{1}+x_{2}$ and $y_{2}$.

A set of alternatives in Proposition~\ref{prop:pMPL3} generalizes the p-MPL decision problem in the third scenario. By rearranging the result to $x_{1}=y_{2}-x_{2}$, we obtain $q=p^{*}$ in the quality elicitation problem. In particular, this shows that the linearity assumption allows us to disentangle the initial endowment and price. To summarize, Proposition~\ref{prop:pMPL3} implies that the injective and the linearity assumptions guarantee that p-MPL accurately elicits quality when subjects combine the initial endowment and price, which we record in Corollary~\ref{cor:pMPL3}.

\begin{cor}\label{cor:pMPL3}
Suppose that a subject chooses according to a weighted separable attribute model of (\ref{model:weighted}) satisfying \ref{assm:injective} (Injective) and \ref{assm:linear} (Linearity). If the subject combines the initial endowment and price in a single attribute so considers quality and earning attributes, then p-MPL accurately elicits quality. Formally, 
\begin{align*}
    V((q,E-p^{*})|\{(q,E-p^{*}),(0,E)\}) &= V((0,E)|\{(0,E),(q,E-p^{*})\})\\
    \text{ if and only if } q&=p^{*}.
\end{align*}
\end{cor}

\begin{table}
\caption{Summary of the required assumptions for accurate quality elicitation}
\begin{center}
\begin{tabular}{lccc}
\toprule
% & Corollary~\ref{cor:mMPL_general} & Corollary~\ref{cor:pMPL1_general} & Corollary~\ref{cor:pMPL2_general}\\ \hline
MPL type & m-MPL & p-MPL & p-MPL \\ 
\text{[Scenario]} & - & \text{[Ignore or Separate]} & \text{[Combine]} \\ 
 \midrule
Assumption & \ref{assm:injective} \text{(Injective)} & \begin{tabular}[c]{@{}c@{}}\ref{assm:injective} \text{(Injective)}\\ \ref{assm:sym} \text{(Symmetry)}\end{tabular} & \begin{tabular}[c]{@{}c@{}}\ref{assm:injective} \text{(Injective)}\\ \ref{assm:linear} \text{(Linearity)}\end{tabular}\\ 
%\midrule
%\text{Domain} & \text{Positive} & \text{Positive \& Negative} & \text{Positive} \\
\bottomrule
\end{tabular}
\end{center}
\vspace{3mm}
\footnotesize{\textit{Notes:} Scenario denotes how subjects treat the initial endowment during the decision-making process. The results are summarized based on Corollaries~\ref{cor:mMPL}, \ref{cor:pMPL12}, and \ref{cor:pMPL3}.}
\label{tab:cor}
\end{table}

Table~\ref{tab:cor} summarizes the required assumptions on a class of weighted separable attribute models for accurate quality elicitation based on Corollaries~\ref{cor:mMPL}, \ref{cor:pMPL12}, and \ref{cor:pMPL3}. There are a few features worth it to note from the table. First, the ignore and the separate scenarios in p-MPL provide the same result. This is because the model we study is additively separable. Second, m-MPL requires fewer assumptions on the models to accurately elicit quality compared to p-MPL. Third, the set of assumptions required for p-MPL to accurately elicit quality depends on how subjects treat the initial endowment. This means that we may need additional assumptions on how subjects treat the initial endowment to guarantee accurate quality elicitation with p-MPL unless an experimenter can control this. In contrast, we do not need such additional assumptions to use m-MPL. 

The second and third points above imply that m-MPL accurately elicits quality for a \textit{wider} range of multi-attribute choice models with \textit{fewer} assumptions compared to p-MPL. To demonstrate the advantage of using m-MPL for accurate quality elicitation, we revisit the range normalization models introduced in Examples~\ref{ex:RN_kinked} and \ref{ex:RN_power}.

For the linear utility case (e.g., $\lambda=1$ in Example~\ref{ex:RN_kinked}), all the assumptions are satisfied. Hence, we conclude that both m-MPL and p-MPL accurately elicit quality given the range normalization model with the linear utility function is true. 

For the kinked utility case as in Example~\ref{ex:RN_kinked}, we can apply Corollaries~\ref{cor:mMPL} and \ref{cor:pMPL3} since the injective and the linearity assumptions are satisfied. However, the symmetry assumption is not satisfied as shown in the example. Thus, Corollary~\ref{cor:pMPL12} is not applicable. Given the range normalization model with the kinked utility function, we conclude that m-MPL guarantees accurate quality elicitation, and p-MPL accurately elicits quality only when subjects combine the initial endowment and price.

For the power utility case as in Example~\ref{ex:RN_power}, Corollaries~\ref{cor:mMPL} and \ref{cor:pMPL12} are applicable since the injective and the symmetry assumptions are satisfied. Yet, we cannot apply Corollary~\ref{cor:pMPL3} since the linearity assumption is not satisfied as shown in the example. Given the range normalization model with the power utility function, we conclude that m-MPL guarantees accurate quality elicitation, and p-MPL accurately elicits quality only when subjects ignore the initial endowment or separate it from price.

\begin{table}
\caption{Quality elicitation results for the range normalization models}
\begin{center}
\begin{tabular}{lccc}
\toprule
MPL method & m-MPL & p-MPL & p-MPL \\
\text{[Scenario]} & - & \text{[Ignore or Separate]} & \text{[Combine]} \\
\midrule
RN with Linear & $\checkmark$ & $\checkmark$ & $\checkmark$ \\
RN with Kinked & $\checkmark$ & $\times$ ($\downarrow$) & $\checkmark$ \\
RN with Power  & $\checkmark$ & $\checkmark$ & $\times$ ($\downarrow$ or $\uparrow$)  \\ 
\bottomrule
\end{tabular}
\end{center}
\vspace{3mm}
\footnotesize{\textit{Notes:} Scenario denotes how subjects treat the initial endowment during the decision-making process. RN denotes the range normalization model introduced in Examples~\ref{ex:RN_kinked} and \ref{ex:RN_power}. The up arrow ($\uparrow$) indicates that the switch point is greater than the model-implied quality. The down arrow ($\downarrow$) indicates that the switch point is lower than the model-implied quality. %\\
%$\dagger$ means that an MPL accurately elicits quality if $\alpha+\gamma\neq0$.
}
\label{tab:RD_sum}
\end{table}

Table~\ref{tab:RD_sum} summarizes the quality elicitation results from the range normalization models. Aligned with the theoretical findings, there are some cases where m-MPL accurately elicits quality, but p-MPL cannot. Moreover, in the kinked utility case, we additionally need to assume that subjects combine the initial endowment with the price to guarantee accurate quality elicitation with p-MPL. In the power utility case, additional assumptions that subjects ignore the initial endowment or separate it from price are required to guarantee accurate quality elicitation with p-MPL. However, m-MPL requires none of such additional assumptions for accurate quality elicitation.

Furthermore, there is no proper assumption on the initial endowment that guarantees accurate quality elicitation with p-MPL when allowing both the kinked and the power utility functions. In contrast, m-MPL can accurately elicit quality even when allowing all types of utility functions mentioned above. In Section~\ref{sec:discuss}, we discuss how the advantages of m-MPL can be put to practical use by connecting it to the experimental study of \cite{somerville2022range}.

\section{Experimental Evidence}\label{sec:experiment}
According to the theoretical findings in Section~\ref{sec:result}, m-MPL requires fewer assumptions compared to p-MPL for accurate quality elicitation. In other words, if p-MPL accurately elicits quality, then m-MPL accurately elicits it as well. Yet, the converse is not true. This implies that the elicited quality values may differ depending on which MPL is used in the experiment. For instance, suppose that a range normalization model with a kinked utility function rationalizes most of the choices. Then according to Example~\ref{ex:RN_kinked}, on average, the quality elicited by p-MPL is lower than the quality elicited by m-MPL.

If the disparity of quality elicitation between the MPLs is negligible, then which MPL to use may not be a huge concern despite the theoretical findings of this paper. However, if the disparity is substantial, then it is informative to check whether there is a systematic difference and how large is the disparity. For exploratory purposes, this section presents the experimental design and evidence to study this question.\footnote{The purpose of the experiment is not to test any specific models nor to conclude which model best describes the data. Instead, the goal is to merely compare the elicitation from the two MPLs.}

\subsection{Experimental Design}
The experiment is designed to compare the quality elicitation by the two MPLs using consumer products. Thirty kinds of snacks were used during the experiment.\footnote{Food is often used as consumer products in experimental studies \citep{sippel1997experiment, wertenbroch2002measuring, anderson2007valuation, krajbich2010visual, huseynov2019distributional}. One benefit of using food as a consumer product in an experiment is that as everyone eats food, there is less concern that subjects are not seriously considering given choice tasks merely because they already own the items \citep{krajbich2012attentional}.
A list of snacks used in this experiment is presented in Appendix~\ref{appendix:experiment}.}
To elicit subjective quality, we find a switch point for either m-MPL or p-MPL. We call \textit{m-block} when m-MPL is used and \textit{p-block} when p-MPL is used.

A within-subject design is used to measure the difference between the quality elicitation by the two MPLs for each subject. This allows us to examine the disparity not only at the aggregate level but also at the individual level. One concern of the within-subject design is that a subject faces an identical set of snacks twice, so the order of the blocks may matter. To control the order effect, the order of the blocks is randomized across sessions. We call \textit{mp-treatment} (\textit{pm-treatment}, respectively) when a subject starts m-block and then p-block (p-block and then m-block, respectively). Additionally, if a subject remembers the choices in the first block, then it may affect the choices in the second block. To mitigate this, the order of the snacks in each block is randomized.

\begin{sidewaysfigure}[!htbp]
\caption{Screenshots of decision pages}\label{fig:screen_decision}
%\caption{Screenshots of decision pages}
\begin{subfigure}[b]{0.49\textwidth}
    \includegraphics[scale=0.3]{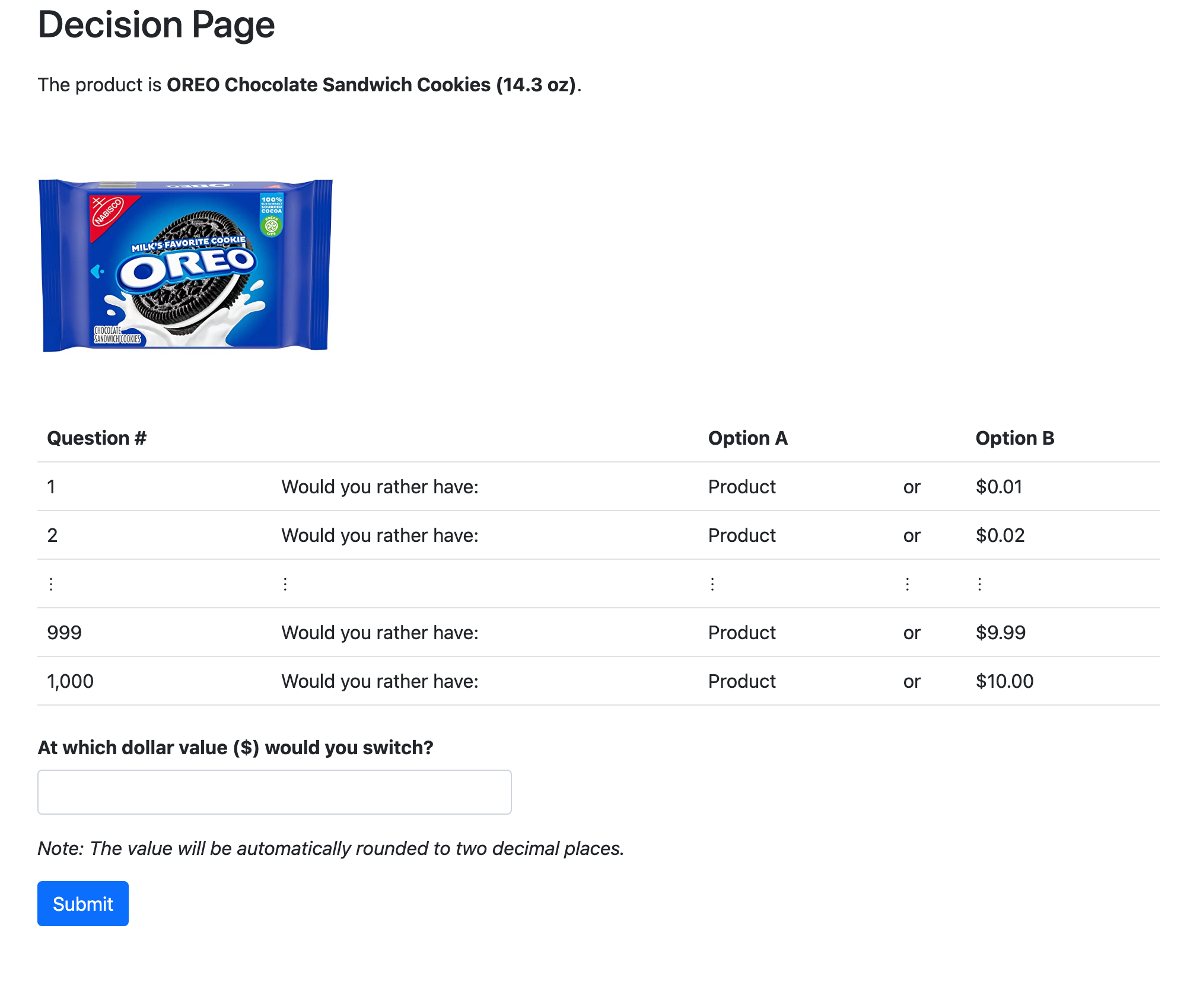}
    \caption{Screenshot of a decision page in m-block}\label{fig:screen_m}
\end{subfigure}
\begin{subfigure}[b]{0.49\textwidth}
    \includegraphics[scale=0.3]{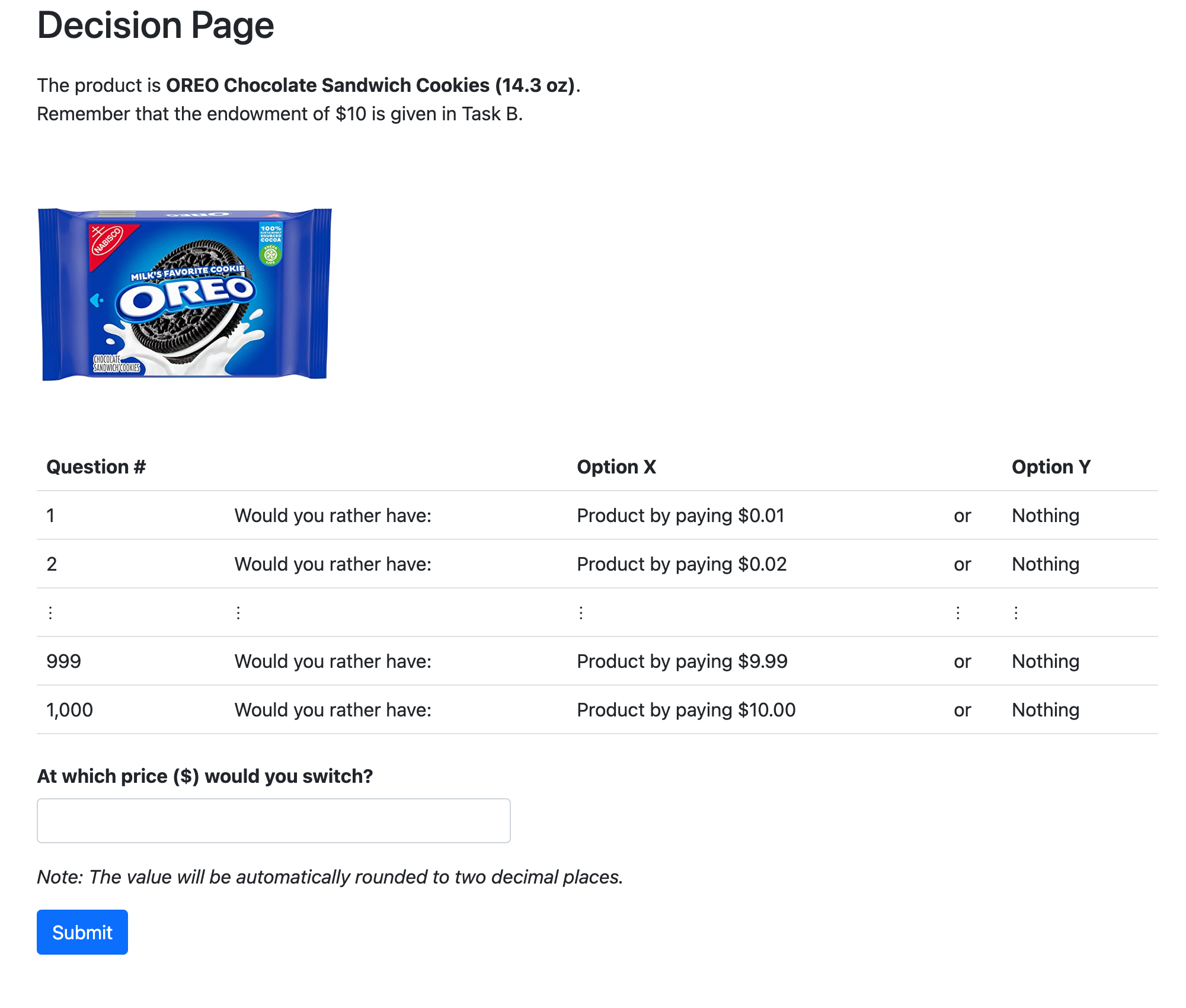}
    \caption{Screenshot of a decision page in p-block}\label{fig:screen_p}
\end{subfigure}
\footnotesize{\textit{Notes:} If a subject reports a switch point, then Option A (or Option X) is chosen from Question 1 to the question right before the switch point, and Option B (or Option Y) is chosen for the remaining questions. The minimum and maximum switch points are 0.01 and 10, respectively, in both blocks. Switch points are rounded to two decimal places.}
\end{sidewaysfigure}

In each round, basic information about a snack such as a picture, name, and weight is presented on a screen. In m-block, subjects also see a set of m-MPL decision problems where Option A is choosing the product and Option B is choosing money with dollar values ranging from \$0.01 to \$10 in increments of one cent (e.g., Figure~\ref{fig:screen_m}). Instead of asking subjects to make 1,000 choices in each round, they are asked at which dollar value they would like to switch from Option A to B. If they report a switch point, then Option A is automatically chosen from Question 1 to the question right before the switch point, and Option B is chosen for the remaining questions. The minimum and maximum switch points are 0.01 and 10, respectively.\footnote{Hence, we are only allowing a single switch point in the experiment. Previous studies show that switching back and forth multiple times in MPLs is often observed \citep{andersen2006elicitation, crosetto2016theoretical, yu2021multiple}. However, we implement a single switch point as the main purpose of the experiment is to compare the switch points from the two MPLs. Roughly speaking, implementing a single switch point can be also understood as assuming strictly increasing normalized weighted utility function, i.e., $u^{0}(t)$ is increasing in $t$.}

In p-block, subjects first receive an initial endowment of \$10. Then they see a set of p-MPL decision problems where Option X is buying the product with prices ranging from \$0.01 to \$10 in increments of one cent and Option Y is doing nothing (e.g., Figure~\ref{fig:screen_p}). Likewise, for each round, subjects are asked to report a switch point where the minimum is 0.01 and the maximum is 10. Once they report a switch point, Option X is chosen for the questions before the switch point, and Option Y is chosen for the questions at and after the switch point. 

The procedure of the experiment is as follows. At the beginning of the experiment, the experimenter reads the general instructions out loud. Then a set of specific instructions for the first block is shown on the screen. Subjects individually read the instructions and solved six comprehension questions. If a subject submitted wrong answers, then the correct answers and explanations are provided. After checking them, they started the main task of the first block. Once they completed 30 switch point decisions, then a new set of specific instructions for the second block was shown on the screen. Again, subjects solved six comprehension questions and then started the task of the second block, reporting 30 switch point decisions.\footnote{The instructions and the screenshots of comprehension questions are presented in Appendix~\ref{appendix:experiment}.}
%Appendix~\ref{appendix:experiment} provides the instructions and the screenshots of comprehension questions.

The payoff was determined by the random problem selection mechanism. Specifically, the computer randomly selected one round and one question after all the tasks were completed. Subjects were paid based on their choices at the selected question of the selected round. A short questionnaire asking about age, gender, race, and major is given at the end of the experiment. After subjects filled out the questionnaire, they received the cash payment in an envelope and a snack when they won it, and left the room.

The experiment was programmed using oTree \citep{chen2016otree}. Subjects were recruited from the Ohio State undergraduate student population through ORSEE \citep{greiner2015subject}. Sessions were conducted in person at the Ohio State University Experimental Economics Laboratory from November to December 2022. 

In total, 85 subjects participated in the experiment, where 42 subjects were in mp-treatment and 43 subjects were in pm-treatment. The average duration was 15 minutes. The average cash payment is \$12.20 including a show-up fee of \$5, and 15 subjects won additional snack items.

\subsection{Data Analysis}
In $b$-block where $b\in\{m,p\}$, subject $i\in\{1,\dots,85\}$ reported a switch point for each product $j\in\{1,\dots,30\}$ based on $b$-MPL. Let $m_{i,j}$ and $p_{i,j}$ be subject $i$'s \textit{product-specific switch points} for product $j$ in m-block and p-block, respectively. We interpret a product-specific switch point as a subjective quality of a specific product elicited by a particular MPL. Each subject reported 30 product-specific switch points in each block. 

Given product $j$, if subject $i$ reported switch points of $0.01$ in both blocks, i.e., $m_{i,j}=p_{i,j}=0.01$, then we assume that the subject dislikes the product. Specifically, let a \textit{nonpositive-value product} be a product in which its product-specific switch points are 0.01 in both m-block and p-block. Let a \textit{positive-value product} be a product that is not a nonpositive-value product. 

Another interpretation of nonpositive-value products is that their subjective quality values are a penny. However, when a nonpositive-value product is observed, we cannot distinguish whether the subject dislikes the product or values it as a penny. Thus, we exclude them from the main analysis, which means that we may have a different number of product-specific switch points for each subject in the main analysis.\footnote{A distribution of the number of nonpositive-value products is presented in Appendix~\ref{appendix:data_cdf}. Around $70\%$ of subjects have 30 positive-value products, and around $90\%$ of subjects have more than 26 positive-value products.}

Let $J_{i,pos}$ be a set collecting all positive-value products for subject $i$. Define subject $i$'s \textit{individual switch points} from m-block and p-block by $m_{i} = \frac{1}{|J_{i,pos}|} \sum_{j\in J_{i,pos}}m_{i,j}$ and $p_{i} = \frac{1}{|J_{i,pos}|} \sum_{j\in J_{i,pos}}p_{i,j}$, respectively. Thus, we interpret an individual switch point as a subject's average subjective quality across products elicited by a particular MPL. Note that only positive-value products are used when computing the individual switch points in the main analysis.\footnote{For robustness check, analysis using the entire dataset including positive-value and nonpositive-value products is provided in Appendix~\ref{appendix:data_entire}. We find that the results are not qualitatively affected by the entire dataset.}

\subsubsection{Individual switch point data}
Prior to the main analysis, we first check whether there is an order effect of blocks using the individual switch point data. To this end, we compare the means of individual switch points between the treatments for each block. Table~\ref{tab:mean} reports the means of individual switch points for each treatment and each block. In m-block, the means are $2.78$ in mp-treatment and $2.95$ in pm-treatment, and they are not significantly different ($p\text{-value}>0.69$, two-sided Wilcoxon rank-sum test). In p-block, the means are $1.81$ in mp-treatment and $1.93$ in pm-treatment, and they are not significantly different as well ($p\text{-value}>0.29$, two-sided Wilcoxon rank-sum test). Thus, no order effect is detected, and we pool the data for all the treatments throughout the analysis.

\begin{table}
\caption{Means of individuals switch points}
\label{tab:mean}
\begin{center}
\begin{tabular}{lcc}
\toprule
             & m-block  & p-block  \\
\midrule
mp-treatment & $2.78$   & $1.81$   \\
             & $(1.15)$ & $(1.51)$ \\
pm-treatment & $2.95$   & $1.93$   \\
             & $(2.08)$ & $(1.08)$ \\

\midrule
Pooled       & $2.87$   & $1.87$   \\
             & $(1.68)$ & $(1.30)$ \\
\bottomrule
\end{tabular}
\end{center}
\vspace{3mm}
    \footnotesize{\textit{Notes:} The table reports the means of individual switch points only using positive-value products. No order effect is detected. Standard deviations are in parentheses.}
\end{table}

Now, we examine the representativeness of the MPLs using the individual switch point data. If both m-MPL and p-MPL accurately elicit quality, then the difference of the individual switch point between the blocks would be close to zero. If the disparity is observed, either positive or negative, then it supports that the quality elicitation using the two MPLs differ, which is suggested by the theoretical findings in Section~\ref{sec:result}. 

The last row in Table~\ref{tab:mean} reports the means of individual switch points from the pooled data. The means are 2.87 in m-block and 1.87 in p-block, and they are significantly different ($p\text{-values}<0.0001$, two-sided paired $t$-test and Wilcoxon signed-rank test).\footnote{There are two subjects whose differences in the individual switch points are greater than or equal to 9. The result is robust even when excluding these outliers. Specifically, the mean individual switch points excluding the outliers are 2.70 in m-block and 1.90 in p-block, and the difference is significantly different ($p$-values$<0.0001$, two-sided paired $t$-test and Wilcoxon signed-rank test).} The result is robust even when the entire dataset including positive-value and nonpositive-value products is considered (see Appendix~\ref{appendix:data_entire}). As a result, our data support the possibility that the quality elicitation by m-MPL and p-MPL are different at the aggregate level. In particular, assuming that m-MPL accurately elicits subjective quality, we can interpret that p-MPL underestimates the quality by $35\%$ according to our data.

\begin{figure}
\caption{CDF of the difference between the individual switch points}
\begin{center}
    \includegraphics[scale=0.45]{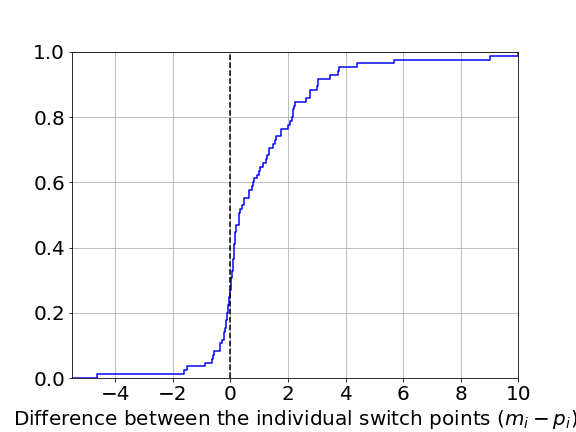}
\end{center}
\footnotesize{\textit{Notes:} A positive (negative, respectively) difference indicates that the subject reported lower (higher, respectively) switch points in p-block compared to m-block, on average. A zero difference indicates that the subject reported the same switch points in m-block and p-block, on average. 
%No subject in our data report zero difference. There are two subjects whose differences are greater than or equal to 9.
}
\label{fig:CDF_diff}
\end{figure}

Figure~\ref{fig:CDF_diff} plots a cumulative distribution function (CDF) of the difference between the individual switch points. This allows us to examine the disparity at the individual level. The figure shows that heterogeneity across subjects exists in the dataset. However, there is a tendency that a majority of subjects (62 out of 85) reported a lower individual switch point in p-block compared to m-block. Using the two-sided sign test, we find that the difference from the median subject is significantly different from zero ($p\text{-value}<0.0001$).\footnote{Details of the sign test are provided in Appendix~\ref{appendix:sign}. The result is robust even when the normal approximation to the binomial distribution is used ($p$-value$<0.0001$) \citep{siegel1988nonparametric}.} The following summarizes the findings from the individual switch point dataset.

\noindent\textbf{Result 1.} Individual switch point data show that, on average, the subjective quality elicited by p-MPL is significantly lower than the subjective quality elicited by m-MPL. While heterogeneity across subjects is detected in the dataset, more than 70\% of subjects reported lower quality values when elicited by p-MPL than by m-MPL.

\subsubsection{Product-specific switch point data}
We use the product-specific switch point data to further examine the heterogeneity across subjects. Recall that in the main analysis, switch points in each block are fewer than 30 for some subjects as nonpositive-value products are excluded. Yet, the dataset is rich enough to statistically test whether a subject behaves differently from a random subject who makes random choices. This allows us to classify subjects into several types and statistically check the heterogeneity in the data.

Define subject $i$'s \textit{absolute m-score} by the number of products in which $m_{i,j}>p_{i,j}$ and \textit{absolute p-score} by the number of products in which $p_{i,j}>m_{i,j}$. Using the binomial distribution with a success rate of $50\%$, we can compute a threshold score where we can statistically say that a subject with an absolute score greater than or equal to the threshold behaves differently from random choices at a 5\% significance level.\footnote{Details on how to compute the threshold are provided in Appendix~\ref{appendix:sign}.} We say that a subject is \textit{m-high} (\textit{p-high}, respectively) type when the subject's absolute m-score (p-score, respectively) is greater than or equal to their threshold score. For example, consider a subject having 30 positive-value products with an m-score of 25 and a p-score of 5. In this case, the threshold is 21, and we classify the subject as m-high type. If a subject has 30 positive-value products with an m-score of 16 and a p-score of 14, then none of the scores passes the threshold of 21. In this case, we classify the subject into neither of the types.

\begin{figure}[t!]
    \caption{CDFs of the normalized scores}
    \begin{center}
        \includegraphics[scale=0.45]{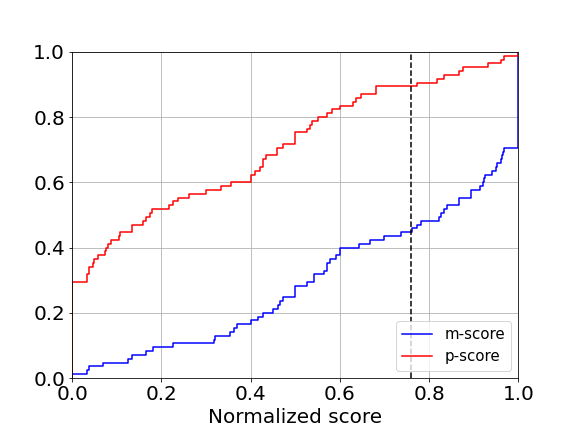}
    \end{center}
    \footnotesize{\textit{Notes:} The vertical dotted line indicates the weighted average threshold of 0.76. Although each subject has a different threshold, the weighted average threshold is still informative as most thresholds are near 0.7 on a percentage scale. 
    %When analyzing the types, one subject who reported 28 nonpositive-value products is additionally excluded since the threshold cannot be computed if the number of positive-value products is less than 9.
    }
    \label{fig:CDFs_scores}
    %\small{Note: The vertical dotted line indicates the weighted average threshold of 0.7143.}
\end{figure}

To compare the scores on an identical scale, we normalize them on a unit interval. Specifically, a subject's \textit{normalized m-score} (\textit{normalized p-score}, respectively) is computed by the subject's absolute m-score (absolute p-score, respectively) divided by the subject's total number of positive-value products. 
Figure~\ref{fig:CDFs_scores} plots the CDFs of the normalized scores. The vertical dotted line in the figure indicates the weighted average threshold of $0.76$. We find that $53\%$ of subjects are classified as m-high type whereas $9\%$ of subjects are classified as p-high type. The other subjects cannot be distinguished at the 5\% significance level. 

Given subject $i$, we say that product $j$ is an \textit{equal-value product} when it is a positive-value product \textit{and} has equal product-specific switch points in both blocks, i.e., $m_{i,j}=p_{i,j}>0.01$. Following a conventional approach, equal-value products are discarded when calculating the scores and thresholds in the above analysis. There are several ways to take equal-value products into account. In this case, each subject's scores and threshold could change. We follow two approaches conducted by \cite{mcgranaghan2023distinguishing} to obtain some bounds on our results.

The first approach is to equally distribute the number of equal-value products to the m-score and p-score. This would make subjects less likely to pass the threshold and provide a lower bound for each type. Using the first approach, $48\%$ of subjects are classified as m-high type whereas $7\%$ are classified as p-high type. The second approach is to proportionally distribute the number of equal-value products to the m-score and p-score based on the ratio of the scores. This would make subjects more likely to pass the threshold and provide an upper bound for each type. Using the second approach, $56\%$ of subjects are classified as m-high type whereas $11\%$ are classified as p-high type.\footnote{A CDF of the number of equal-value products and CDFs of the scores incorporating equal-value products are presented in Appendix~\ref{appendix:data_cdf}.} The following summarizes the results of classifying types.

\noindent\textbf{Result 2.} Product-specific switch point data show that $48-56\%$ of subjects are classified as m-high type and $7-11\%$ of subjects are classified as p-high type at the 5\% significance level. 

Using the product-specific switch point data, we can run regressions to examine whether the disparity between the switch points is significant while controlling for other variables and fixed effects. Let $Switch_{i,j,b}$ be subject $i$'s switch point for product $j$ in $b$-block and $Block_{b}$ be an indicator variable in which $Block_{m}=0$ and $Block_{p}=1$. Our primary interest is to regress $Switch$ on $Block$ since the coefficient of $Block$ indicates the disparity between the quality elicitation by the two MPLs. 

\begin{table}[t!]
    \caption{Regression results}\label{tab:reg}
        \begin{center}

%            \resizebox{1\textwidth}{!}{
%            \input{Attribute/Tables/Regression.tex}
%}
{
\def\sym#1{\ifmmode^{#1}\else\(^{#1}\)\fi}
%\begin{tabular}{l*{4}{D{.}{.}{-1}}}
\begin{tabular}{lccc}
\toprule
                    & (1)              & (2)              & (3)                          \\
\midrule
Block               & $-0.988\sym{***}$& $-0.988\sym{***}$& $-0.988\sym{***}$ \\
                    & $(0.229)$        & $(0.231)$        & $(0.232)$         \\
\addlinespace
Price               & $0.286\sym{***}$ & $0.290\sym{***}$ & -                 \\
                    & $(0.029)$        & $(0.029)$        & -                 \\
\addlinespace
Block$\times$Price  & $-0.003$         & $-0.003$         & $-0.003$          \\
                    & $(0.024)$        & $(0.024)$        & $(0.024)$         \\
\addlinespace
Constant            & $2.181\sym{***}$ & $1.155\sym{***}$ & $2.254\sym{***}$  \\
                    & $(0.207)$        & $(0.151)$        & $(0.198)$         \\
\midrule
Individual fixed effects & No          & Yes              &  Yes              \\
Product fixed effects    & No          & No               &  Yes              \\
\midrule     
Observations        & $4,798$          & $4,798$          & $4,798$           \\
%Left-censored       & -                &-                 \\
%Right-censored      & -                &-                 \\
Number of subjects  & $85$             & $85$             & $85$              \\
\bottomrule
\end{tabular}
}

\end{center}
\vspace{3mm}
\footnotesize{\textit{Notes:} The dependent variable is $Switch_{i,j,b}$ denoting subject $i$'s switch point for product $j$ in $b$-block. $Block_{b}$ denotes a block indicator where $Block_{m}=0$ and $Block_{p}=1$. $Price_{j}$ denotes the market price of product $j$. A coefficient of $Price$ is omitted when the product fixed effects are included because a market price is invariant for each product.
Clustered standard errors at the individual level are in parentheses. \\ * $p<0.10$, ** $p<0.05$, *** $p<0.01$.}
\end{table}

One may be concerned about different price sensitivity between the blocks due to an anchor effect. More specifically, while subjects are asked to choose between a product and money in m-MPL, p-MPL asks subjects whether they would like to buy a product or not. The p-MPL form of questioning can lead subjects to want to sell the product after the experiment. If subjects think about re-selling during the decision-making process in p-block, then they would be more sensitive to market price when p-MPL is used. Let $Price_{j}$ be the market price of product $j$. To control for the anchor effect, we include $Price$ and the interaction term of $Price$ and $Block$.\footnote{Rigorously speaking, subjects' beliefs on prices should be considered to control for the anchor effect. Yet, beliefs on prices are not elicited during the experiment. Thus, the market prices are used as a proxy.} 

Table~\ref{tab:reg} reports the regression results from the linear model. Column 1 reports the result without individual and product fixed effects, column 2 reports the result including the individual fixed effects, and column 3 reports the result including both the individual and product fixed effects. 

First, all the estimated coefficients of $Block$ are negative and significantly different from zero. This implies that the subjective quality elicited by p-MPL is lower than the quality by m-MPL. The result is robust even when considering the entire dataset (see Appendix~\ref{appendix:data_entire}). 

Next, we check whether the anchor effect exists using the regression results. The coefficients of $Price$ in columns 1 and 2 are all positive and significant, indicating that subjects are sensitive to the price when making switch point decisions. However, note that the coefficient is less than one so the quality measure is not driven entirely by price.  Meanwhile, the coefficients of $Block \times Price$ in columns 1, 2, and 3 are close to zero and insignificant, which implies that the price sensitivity does not depend on the MPL used in the experiment. Thus, we find no evidence that p-block particularly draws attention to the market price that differs from m-block.\footnote{While we find no evidence of distinct price sensitivities between the two blocks, it is still possible that the anchor effect or a bias toward the market price equally occurs in both MPLs. In this paper, the representation of the MPL decision problems abstracts from other behavioral biases such as rounding and left-digit biases. How to formally incorporate them in the MPL representation and whether they create a systematic bias are open questions.} The following summarizes the findings from the regression results.

\noindent\textbf{Result 3.} The regression results show that the quality elicited by p-MPL is significantly lower than the quality elicited by m-MPL, which is consistent with Results 1 and 2. While subjects are sensitive to prices when making decisions, price sensitivity does not differ between the MPLs.

Given that m-MPL requires fewer assumptions compared to p-MPL for accurate quality elicitation and the disparity observed in our experimental data, one may conclude that p-MPL is not suited for measuring quality. However, we need to be cautious in concluding that p-MPL is obsolete in a multi-attribute choice setting. The decision problem in p-MPL is asking a subject how much they are willing to buy a product, which is a measure of willingness to pay (WTP).\footnote{\cite{cunningham2013comparisons} theoretically examines WTP in a multi-attribute choice setting under general contexts. A switch point from p-MPL in this paper aligns with the WTP definition by \cite{cunningham2013comparisons}.} By interpreting that m-MPL measures quality and p-MPL measures WTP, our experimental data show that their values differ. Thus, our findings suggest that a researcher may have to choose which MPL to use depending on the research question. More discussion on the experimental findings is provided in Section~\ref{sec:discuss}.

\section{Discussion}\label{sec:discuss}
In this section, we discuss the theoretical and experimental findings of this paper by relating them to previous literature. First, we link our experimental findings to the endowment effect studied by \cite{kahneman1990experimental}. We further discuss possible rationales for the disparity observed in our experiment. Second, we clarify the theoretical framework of an upgrading method proposed by \cite{park2008eliciting}. Third, we demonstrate the advantage of using m-MPL when testing multi-attribute choice models following the experimental design of \cite{somerville2022range}.

\subsection{Interpretations of the experimental findings}\label{sec:discuss_data}
A seminal paper by \cite{kahneman1990experimental} experimentally studies the disparity between the willingness to accept (WTA) and the willingness to pay (WTP), also known as an endowment effect. In their between-subject experiment, they decompose the disparity into two parts: \textit{reluctance to buy} and \textit{reluctance to sell}. Specifically, subjects are assigned the role of either buyer, chooser, or seller. Buyers were asked whether they would like to buy a mug at a price ranging from \$0 to \$9.25 or not. Choosers were asked whether they would like to choose a mug or money, where the monetary value is also ranging from \$0 to \$9.25. Sellers received a mug and were asked whether they would like to sell it at each of the possible prices or keep it.

Using an MPL-type method, they elicit the WTA from sellers and the WTP from buyers. As the choices from choosers are neutral, the degree of reluctance to buy is measured by the gap between the elicited values from buyers and choosers. The degree of reluctance to sell is measured by the gap between the elicited values from sellers and choosers. Their data show that the median valuations are \$2.87, \$3.12, and \$7.12 for buyers, choosers, and sellers, respectively. They run another similar experiment using mugs and find that the median valuations are \$2, \$3.50, and \$7 for buyers, choosers, and sellers, respectively. Therefore, they find evidence of both the reluctance to buy and the reluctance to sell.

Based on \cite{kahneman1990experimental}, we can understand the disparity observed in our data as reluctance to buy. The reason is that the decision problems of choosers and buyers are similar to those in m-block and p-block, respectively. Thus, our data from the within-subject experiment support that reluctance to buy exists for a variety of consumer products. While \cite{kahneman1990experimental} find evidence of reluctance to buy, its statistical significance is inconclusive since no statistical test results are reported. The analysis in this paper complements this. The statistical results presented in Section~\ref{sec:experiment} support that the size of the reluctance to buy is significant.

A close study investigating the difference between quality and other measures in a multi-attribute choice setting is \cite{bordalo2012saliencein}. They distinguish an endowment stage where a decision maker evaluates a product by thinking that they are endowed with it and a trading stage where a decision maker evaluates a product by thinking about how much to accept when they sell it. They theoretically show that the WTA can exceed quality due to an intrinsic value of ownership from the endowment stage. In this case, the disparity is related to reluctance to sell. In our experiment, there is no step that can be considered as the endowment stage in m-block. The instructions also never mentioned that the products are endowed. Hence, the disparity observed in our experiment is distinct from the interpretation suggested by \cite{bordalo2012saliencein}.

Another possible interpretation for the disparity is a wealth effect. Recall that there is no initial endowment in m-block, whereas subjects receive \$10 before making decisions in p-block. Our theoretical results imply that if the required assumptions are satisfied, then there should be no wealth effect and both MPLs accurately elicit quality. However, we find a disparity in our dataset, which suggests that some assumptions are violated and wealth may play a role. 
%Our experimental design cannot capture how wealth affects the disparity, and a future study examining the relationship between them would be interesting.

A framing effect could be involved in the disparity as well. The way subjects treat an initial endowment may depend on how it is presented during the experiment. On a decision page in p-block, we clearly mentioned that the initial endowment of \$10 is given to subjects (see Figure~\ref{fig:screen_p}). We may have different results when the initial endowment is presented in different ways. For example, we can inform subjects that \$10 is initially given in the instructions but omit the reminder sentence on a decision page. Exploring the framing effect would be informative since it can provide useful information such as to what extent an experimenter can control by design how subjects encode attributes.\footnote{\cite{kHoszegi2013model} discuss how framing can affect the representation of alternatives in a multi-attribute choice setting. \cite{dertwinkel2022concentration} experimentally show that a concentration bias in intertemporal choices is influenced by how the costs and payoffs of a task are presented on a decision page.}

We briefly discuss how different assumptions on how the show-up fee enters the model affect the results. Throughout the paper, we assume that subjects ignore the show-up fee or separate it from money and price.\footnote{Let $S$ be a show-up fee. If the show-up fee is ignored, then it is obvious that all the results in Section~\ref{sec:result} hold. Suppose that the show-up fee is separated from the money and price attributes. Then one possible representation would be $x=(q,0,S)$ and $y=(0,m,S)$ in m-MPL and $x=(q,-p,S)$ and $y=(0,0,S)$ in p-MPL. As long as the show-up fee is separated from money and price, all the theoretical results hold due to the additively separable model (see Propositions~\ref{prop:mMPL}, \ref{prop:pMPL12}, and \ref{prop:pMPL3}).} We believe that this is a reasonable assumption because the show-up fee does not interact with the decision problems. In detail, the information about the show-up fee is introduced at the beginning of the experiment and never mentioned again until the payment stage. Still, one could think that subjects consider the show-up fee and combine it with money and price.\footnote{For example, one possible representation would be $x=(q,S)$ and $y=(0,m+S)$ in m-MPL and $x=(q,S-p)$ and $y=(0,S)$ in p-MPL in which $S$ denotes a show-up fee.} In this case, Proposition~\ref{prop:pMPL3} implies that both m-MPL and p-MPL require the injective and the linearity assumptions to accurately elicit quality. This suggests that if the linearity assumption is violated, then both MPLs cannot accurately elicit quality, which may create the disparity. How subjects treat a show-up fee in an experiment could be related to the framing effect discussed above.

Lastly, our experiment is not designed to test models nor investigate which behavioral factors affect the elicitation. However, we can still make conjectures about which behavioral factor causes the disparity. As illustrated in Example~\ref{ex:RN_kinked}, loss aversion is one candidate since p-MPL underestimates quality given the range normalization model with a kinked utility function. As shown in Example~\ref{ex:RN_power}, increasing marginal utility is another candidate since p-MPL underestimates quality given the range normalization model with a convex utility function.\footnote{Example~\ref{ex:PN} in Appendix~\ref{appendix:example} also suggests a possibility of the underestimation of quality when p-MPL is used given a pairwise normalization model with the power utility function.} Given the results of this paper and recent findings of a low correlation between loss aversion and the endowment effect \citep{chapman2023willingness, fehr2022endowment}, it might be interesting to study how other behavioral factors such as increasing marginal utility are correlated with the endowment effect.

\subsection{Attribute value elicitation for finitely many attributes}\label{sec:discuss_update}
A marketing paper by \cite{park2008eliciting} proposes an upgrading method to measure attribute values for products with finitely many attributes. The basic idea is that, even when there are many attributes, we can measure the \textit{marginal} attribute value of an attribute by solely upgrading that single attribute. 

To illustrate, suppose that a researcher wants to study consumer behavior on cars, and the attributes of interest are fuel economy, safety, comfort, and navigation system. Consider two completely identical vehicles except for the fuel economy attribute.\footnote{For instance, the first vehicle requires an average of 8.1 liters of super unleaded per 100 km and the second vehicle requires an average of 9.5 liters of super unleaded per 100 km. This example is borrowed from the supplementary appendix in \cite{mrkva2020moderating}.} To measure the marginal fuel economy attribute value or the difference in the fuel economy attribute values, the researcher offers two alternatives. One is choosing the superior vehicle and the other one is choosing the inferior vehicle plus a bonus denoted by $d$. Importantly, the researcher asks a decision maker how much bonus is needed to make the two alternatives indifferent.\footnote{The bonus can be understood as the minimum willingness to accept to change from a superior product to an inferior product. The original design of the upgrading method proposed by \cite{park2008eliciting} asks subjects to report the maximum willingness to pay to change from an inferior product to a superior product.} 

Formally, let $q_{1}$, $q_{2}$, $q_{3}$, and $q_{4}$ be attribute values of the fuel economy, safety, comfort, and navigation system attributes expressed in dollars, respectively, related to the superior vehicle. Let $\tilde{q}_{1}$ be the difference in the fuel economy attribute values between the two vehicles. In a multi-attribute choice setting, the first alternative is written as $(q_{1},q_{2},q_{3},q_{4},0)$ and the second alternative is written as $(q_{1}-\tilde{q}_{1},q_{2},q_{3},q_{4},d)$ where the fifth attribute is associated with a bonus attribute.

Suppose that a decision maker reports $d^{*}$. Then Proposition~\ref{prop:pMPL3} implies that $V((q_{1},q_{2},q_{3},q_{4},0)|\cdot)=V((q_{1}-\tilde{q}_{1},q_{2},q_{3},q_{4},d^{*})|\cdot)$ is equivalent to $\tilde{q}_{1}=d^{*}$ for any weighted separable attribute model. Hence, when the injective and the linearity assumptions hold, the marginal fuel economy attribute value is accurately elicited. Similarly, other attribute values can be accurately elicited by upgrading the product one attribute by one attribute.

In fact, all the decision problems studied in the propositions resemble choices having balanced trade-offs where one alternative has the same number of advantageous and disadvantageous attributes relative to the other alternative \citep{kHoszegi2013model}.\footnote{For example, when $x=(x_{1},\dots,x_{N})$ and $y=(y_{1},\dots,y_{N})$ in which $x_{1}>y_{1}$, $y_{2}>x_{2}$, and $x_{n}=y_{n}$ for all $n\in \{3,\dots,N\}$ are given, then $x$ and $y$ have balanced trade-offs with one advantageous attribute and one disadvantageous attribute.} \cite{dertwinkel2022concentration} show that a switch point decision for balanced trade-off alternatives is not influenced by a bias toward concentration in intertemporal choices for a range normalization model. Similarly, we can understand the theoretical results of this paper as the switch point decision is not distorted by the weights presented in any weighted separable attribute model.\footnote{Specifically, consider $x=(x_{1},\dots,x_{N})$ and $y=(y_{1},\dots,y_{N})$ from Propositions~\ref{prop:mMPL}, \ref{prop:pMPL12}, or \ref{prop:pMPL3}. Then we have $\sum_{n}x_{n} = \sum_{n}y_{n}$ at the switch point when proper assumptions are satisfied.}

\subsection{Testing multi-attribute choice models}\label{sec:discuss_somerville}
\cite{somerville2022range} proposes an experimental design to test multi-attribute choice models in a consumer context. The experiment consists of two parts: an elicitation step and a main choice task. In the elicitation step, subjective quality values of products are elicited. They are used to attain prediction regions as explained below. In the main choice task, there are three alternatives: high-quality product, low-quality product, and outside option denoted by $h$, $\ell$, and $o$, respectively. Let $h_{q}$ and $\ell_{q}$ be the subjective quality values of high-quality and low-quality products, respectively. Let $h_{p}$ and $\ell_{p}$ be the prices of high-quality and low-quality products, respectively. Assume that a subjective quality value and price are zero for an outside option. 

\begin{figure}[t!]
\caption{Example of prediction regions by \cite{somerville2022range}}\label{fig:somervillen13}
\begin{center}
    \includegraphics[scale=0.5]{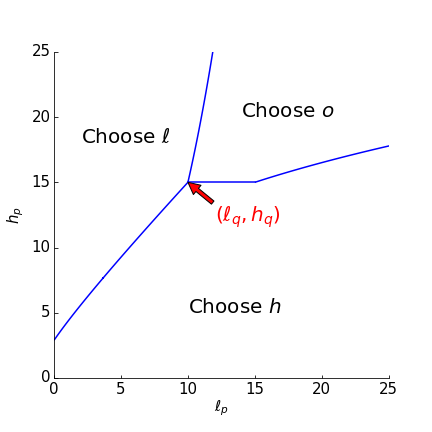}
\end{center}
    \footnotesize{\textit{Notes:} Blue lines indicate indifference curves. Prediction regions are pinned down by the elicited quality values of $\ell_{q}=10$ and $h_{q}=15$.}
\end{figure}

To illustrate how to attain prediction regions, consider the range normalization model in Example~\ref{ex:RN_kinked} with $\gamma=-\frac{1}{3}$ and $\lambda=1$ (i.e., linear utility). Suppose that the elicited quality values from the elicitation step are $\ell_{q} = 10$ and $h_{q}=15$. Then we can draw indifference curves on a price domain as depicted in Figure~\ref{fig:somervillen13}.\footnote{Each decision problem in the main choice task consists of three alternatives. In this case, the model has a weight function such as $w_{RN}(t,s,r) = |\max\{u(t),u(s),u(r)\}-\min\{u(t),u(s),u(r)\}|^{\gamma}$.} As shown in the figure, we obtain complete prediction regions that are pinned down by the elicited values. For example, in the main choice task, if prices are given as $\ell_{p}=10$ and $h_{p}=20$, then the model predicts that a low-quality product would be chosen. We can examine how the model performs by comparing the predictions to the actual choices from a series of decision problems in the main choice task. 

\cite{somerville2022range} tests the range normalization models with the linear utility function. In their experiment, the p-MPL type method is used to elicit quality. Our theoretical results show that both m-MPL and p-MPL guarantee accurate quality elicitation for the linear utility case as summarized in Table~\ref{tab:RD_sum}. Thus, using p-MPL is not an issue from a theoretical point of view. However, as our experimental data suggest, the quality elicitation by m-MPL and p-MPL could differ in the context of \cite{somerville2022range}. If the disparity is present, then caution should be exercised when using p-MPL. For example, prediction regions for the range normalization model with the linear utility function might be misspecified as p-MPL underestimates quality. If there is noise in the prediction regions, then choices might not accurately separate different models.\footnote{Whether the disparity occurs when using different MPLs and how it affects the conclusion in \cite{somerville2022range} are open questions. For instance, by estimating $\gamma$ assuming heterogeneous-agent, \cite{somerville2022range} finds that 75\% of subjects are consistent with $\gamma<0$. The fraction of those subjects could change when m-MPL is used. However, the fraction may increase, decrease, or depend on the composition of the subjects' types.}

Furthermore, suppose that one wants to extend \cite{somerville2022range} by testing the range normalization models with different types of utility functions such as kinked and/or power utility functions as introduced in Examples~\ref{ex:RN_kinked} and \ref{ex:RN_power}. Then the first step is to check whether accurate quality elicitation is guaranteed for these models. For the kinked and the power utility cases, we need additional assumptions on how subjects treat the initial endowment to accurately elicit quality with p-MPL. In contrast, m-MPL guarantees accurate quality elicitation without such additional assumptions for those models. Moreover, m-MPL can be used even when we want to test all the models at the same time. This implies that a single elicitation step with m-MPL is enough to simultaneously test multiple models. Meanwhile, using p-MPL may introduce systematic noise.

Appendix~\ref{appendix:example} provides more examples using different multi-attribute choice models such as the pairwise normalization models \citep{landry2021pairwise} and the contextual concavity models \citep{kivetz2004alternative}. Examples in the appendix confirm that a wider range of multi-attribute choice models can be simultaneously tested by implementing a single elicitation step with m-MPL.

\section{Conclusion}\label{sec:conclusion}
This paper studies the quality elicitation problem in a multi-attribute choice setting. Theoretical results show that m-MPL can accurately elicit subjective quality for a wider range of multi-attribute choice models with fewer assumptions compared to p-MPL. For exploratory purposes, the paper conducts a within-subject experiment using consumer products and finds that the quality elicited by p-MPL is significantly lower than the quality elicited by m-MPL.

Both theoretical and experimental findings in this paper suggest that p-MPL may provide inaccurate quality elicitation in a multi-attribute choice setting. However, this does not mean that p-MPL is obsolete. As discussed in Sections~\ref{sec:experiment} and \ref{sec:discuss}, we can understand that p-MPL is a measure of WTP rather than quality. Hence, if a researcher is interested in studying quality, then our findings suggest using m-MPL. If the WTP is the main research topic, then using p-MPL seems more appropriate.

Although the paper focuses on quality elicitation for consumer products, the theoretical results can be applied to general contexts. For instance, \cite{je2023does} measures the value of signals using the p-MPL type method. In the experiment, subjects had to guess the color of a drawn ball from a box. The winning probability without a signal is 50\%. However, subjects had a chance to purchase a signal that increased the average winning probability to 70\% before guessing the color. In a multi-attribute choice setting, we can think that there are two attributes of interest: a signal attribute and a price attribute. Corollaries \ref{cor:pMPL12} and \ref{cor:pMPL3} suggest that the symmetry or the linearity assumption in addition to the injective assumption are needed in the underlying model to conclude that the value of the signal is equal to the price at the switch point.

Multi-attribute choices are extensively studied in various fields including marketing, psychology, and neuroscience \citep{huber1982adding, simonson1989choice, netzer2011adaptive, trueblood2013not, mrkva2020moderating, soltani2012range, busemeyer2019cognitive}. Since the approach used in this paper is useful to not only economists but also researchers in other disciplines, we believe that this paper elucidates methodological choices relevant to multiple disciplines.

\newpage

%%%%%%%%%%%%%%%%%%%%%%%%%%%%%%%%%%%%%%%%%%%%%%%%%%%%%%%%%%%%%%%%
%%%%%%%%%%%%%%%%%%%%%%%%% Appendix %%%%%%%%%%%%%%%%%%%%%%%%%%%%%
%%%%%%%%%%%%%%%%%%%%%%%%%%%%%%%%%%%%%%%%%%%%%%%%%%%%%%%%%%%%%%%%

\appendix
\section*{Appendix}
\section{Theorems}\label{appendix:thm}
In this section, the theorems and their proofs used for the main theoretical findings in Section~\ref{sec:result} are provided. We first define a more general class of multi-attribute choice models that nests weighted separable attribute models as defined in (\ref{model:weighted}).

As in Section~\ref{sec:framework}, we focus on a binary choice problem and define an \textit{evaluation function} by $V:\mathbb{R}^{N} \times \mathbb{R}^{N} \rightarrow \mathbb{R}$. Let $v:\mathbb{R} \times \mathbb{R} \rightarrow \mathbb{R}$ be an \textit{attribute evaluation function} that is $v(0,t)=0$ for all $t\in\mathbb{R}$ and bounded so $|v(t,s)|<\infty$ for all $t,s\in \mathbb{R}$. Its value indicates an evaluation of an attribute value where the first argument is the evaluated attribute value and the second argument is the comparable attribute value. For example, $v(x_{n},y_{n})$ is the evaluation of attribute value $x_{n}$ when $y_{n}$ is the comparable attribute value. Thus, the evaluation of an attribute value may depend on attribute values in the same attribute from other alternatives, which allows menu-dependent context effects.

Define a \textit{normalized attribute evaluation function} by $v^{0}(t)=v(t,0)$ for all $t\in\mathbb{R}$. It indicates the evaluation of an attribute value when the context effect is normalized by the zero comparable attribute value. In addition, since $v^{0}(t)=v(t,0)-v(0,t)$, it can be interpreted as \textit{marginal attribute evaluation} that a decision maker attains from an attribute value of $t$ instead of 0 when both attribute values are available. We assume that $v^{0}(t)\geq0$ for $t>0$ and $v^{0}(t)\leq0$ for $t<0$. Now, given a menu with two alternatives, we define a class of multi-attribute choice models in which its evaluation of an alternative is the sum of attribute evaluations.

\begin{defn}[Separable attribute models]\label{class:separable}
Let two alternatives $x=(x_{1},\dots,x_{N})$ and $y=(y_{1},\dots,y_{N})$ in which $x_{n}y_{n} \geq 0$ for all $n\in \{1,\dots,N\}$ be given. A model is called a separable attribute model when its evaluation is the sum of attribute evaluations. Formally,
\begin{equation}\label{model:separable}
    V(x|\{x,y\}) = \sum_{n=1}^{N} v(x_{n},y_{n}).
\end{equation}
\end{defn}

%An attribute evaluation function above normalizes the evaluation of zero attribute value to zero regardless of the other attribute value, i.e., $v(0,t)=0$ for all $t$. This implies that the evaluation of alternative $(0,\dots,0)$ by a separable attribute model is always zero. Fixing the other attribute value equals zero, the attribute evaluation of any positive (negative, respectively) attribute value is nonnegative (nonpositive, respectively), i.e., $v(t,0)\geq0$ for $t>0$ ($v(t,0)\leq0$ for $t<0$, respectively).

Consider any utility function $u:\mathbb{R}\rightarrow\mathbb{R}$ and weight function $w:\mathbb{R}\times\mathbb{R}\rightarrow\mathbb{R}_{+}$ from a weighted separable attribute model of (\ref{model:weighted}). Define $v(t,s)=u(t)w(t,s)$ for all $t,s\in\mathbb{R}$. Since $u(t)$ and $w(t,s)$ are bounded, $v(t,s)$ is also bounded. In addition, since $u(0)=0$, it follows that $v(0,t)=0$. Note that $v^{0}(t)=v(t,0)=u(t)w(t,0)=u^{0}(t)$. Since $u(t)$ is weakly increasing and $w(t,s)$ is nonnegative, we have $v^{0}(t)\geq0$ for $t>0$ and $v^{0}(t)\leq0$ for $t<0$. Therefore, a weighted separable attribute model is nested in a class of separable attribute models.

A class of separable attribute models also nests some multi-attribute choice models with utility functions that depend on a menu. Formally, define a \textit{menu-dependent utility function} by a function $u:\mathbb{R}\times\mathbb{R}\rightarrow\mathbb{R}$ such that $u(0,t)=0$ for all $t\in\mathbb{R}$, weakly increasing with respect to the first argument, and $|u(t,s)|<\infty$ for all $t,s\in\mathbb{R}$. Consider a multi-attribute choice model that evaluates $x$ in $\{x,y\}$ by
\begin{equation}\label{model:weighted_menu}
    V(x|\{x,y\}) = \sum_{n=1}^{N} u(x_{n},y_{n}) w(x_{n},y_{n}),
\end{equation}
where $w:\mathbb{R}\times\mathbb{R}\rightarrow\mathbb{R}_{+}$ is a weight function as defined in a weighted separable attribute model of (\ref{model:weighted}). Define $v(t,s)=u(t,s)w(t,s)$ for all $t,s\in\mathbb{R}$. Then $v(t,s)$ is bounded and $v(0,t)=0$. In addition, we have $v^{0}(t)=u(t,0)w(t,0)\geq0$ for $t>0$ and $v^{0}(t)\leq0$ for $t<0$. Hence, model (\ref{model:weighted_menu}) is a separable attribute model. Examples of model (\ref{model:weighted_menu}) include the contextual concavity models of \cite{kivetz2004alternative}.

Now, to study the quality elicitation problems, we introduce assumptions on a class of separable attribute models.

\begin{assm}
Consider a separable attribute model of (\ref{model:separable}).
\begin{enumerate}[label=B\theenumi]
    \item (Injective) If $t>s>0$, then $v^{0}(t) \neq v^{0}(s)$.\label{assm:injective_general}
    \item (Symmetry) For any $t>0$, we have $v^{0}(t) = -v^{0}(-t)$.\label{assm:sym_general}
    \item (Linearity) For any $t>s>0$, we have $v^{0}(t-s) = v(t,s) - v(s,t)$.\label{assm:linear_general}
\end{enumerate}
\end{assm}

%If the one-to-one assumption holds, then each attribute value has a unique normalized attribute evaluation. The one-to-one assumption and continuity together imply strict monotonicity. The symmetry assumption indicates that the size of a gain from an attribute value and the size of a loss from the same attribute value are identical. In this vein, we can understand the assumption as there is no loss aversion property for each dimension. Lastly, if the linearity assumption holds, then it shows that a normalized attribute evaluation can be linearly decomposed into two attribute evaluations.

The implications of the assumptions discussed in Section~\ref{sec:framework} hold for these assumptions as well. The injective assumption of \ref{assm:injective_general} implies that each positive attribute value has a unique normalized attribute evaluation. The symmetry assumption of \ref{assm:sym_general} implies that gains and losses are symmetric. The linearity assumption of \ref{assm:linear_general} implies constant marginal attribute evaluation whenever the distance between two attribute values is the same.

The theorems and their proofs used for the main results are presented below. Theorem~\ref{thm:mMPL} generalizes Proposition~\ref{prop:mMPL}, which shows that given two-attribute alternatives consisting of one positive attribute value and one zero attribute value, a separable attribute model links distinct attributes one to one under the injective assumption.

\begin{theorem}\label{thm:mMPL}
Let two-attribute alternatives $x=(x_{1},0)$ and $y=(0,y_{2})$ with $x_{1},y_{2}>0$ be given. Suppose that a decision maker chooses according to a separable attribute model of (\ref{model:separable}). Then the following two statements are equivalent.
\begin{enumerate}[label=(\alph*)]
    \item The model satisfies \ref{assm:injective_general} (Injective).
    \item $V(x|\{x,y\}) = V(y|\{y,x\})$ if and only if $x_{1}=y_{2}$.
\end{enumerate}
The result holds even when alternatives have additional attributes with constants, for example, $x=(x_{1},0,c_{3},\dots,c_{N})$ and $y=(0,y_{2},c_{3},\dots,c_{N})$ in which $c_{3},\dots,c_{N} \in \mathbb{R}$.
\end{theorem}

\begin{proof}[Proof of Theorem~\ref{thm:mMPL}]
[(a)$\implies$(b)]
By definition, $V(x|\{x,y\}) = V(y|\{y,x\})$ is equivalent to $v(x_{1},0) = v(y_{2},0)$. This can be rewritten as $v^{0}(x_{1}) = v^{0}(y_{2})$. Due to the injective assumption of \ref{assm:injective_general}, we conclude that $x_{1} = y_{2}$. Next, suppose that $x_{1} = y_{2}$. Then we have $v^{0}(x_{1}) = v^{0}(y_{2})$, which implies that $V(x|\{x,y\}) = V(y|\{y,x\})$.

[(b)$\implies$(a)] 
Let $t>s>0$ be given. Consider $x=(t,0)$ and $y=(0,s)$. By (b), it follows that $v^{0}(t) = V(x|\{x,y\}) \neq V(y|\{y,x\}) = v^{0}(s)$.

The same logic can be applied to prove the case when the alternatives have additional attributes with as the model is additively separable. 
%Specifically, whenever additional attribute with constant $c\in\mathbb{R}$ is added to both alternatives, any separable attribute model adds an equal value of $v(c,c)$ to both evaluations of alternatives.
\end{proof}

As discussed in Section~\ref{sec:result}, a set of alternatives above generalizes the m-MPL decision problem. Thus, the result indicates $q=m^{*}$ in the quality elicitation problem. Corollary~\ref{cor:mMPL_general} records the implication of Theorem~\ref{thm:mMPL} that the injective assumption alone guarantees that m-MPL accurately elicits quality for any separable attribute model.

\begin{cor}\label{cor:mMPL_general}
Suppose that a subject chooses according to a separable attribute model of (\ref{model:separable}) satisfying \ref{assm:injective_general} (Injective). If the subject considers quality and money attributes, then m-MPL accurately elicits quality. Formally,
    \begin{align*}
        V((q,0)|\{(q,0),(0,m^{*})\}) &= V((0,m^{*})|\{(0,m^{*}),(q,0)\}) \\
        \text{ if and only if } q&=m^{*}.
    \end{align*}
\end{cor}

Next, we examine the quality elicitation problem for the p-MPL cases. Theorem~\ref{thm:pMPL12} generalizes Proposition~\ref{prop:pMPL12} which shows that given two-attribute alternatives where one of them consists of zero attribute values, a separable attribute model links distinct attributes one to one under the injective and the symmetry assumptions.

\begin{theorem}\label{thm:pMPL12}
Let two-attribute alternatives $x=(x_{1},x_{2})$ and $y=(0,0)$ with $x_{1}x_{2}<0$ be given. Suppose that a decision maker chooses according to a separable attribute model of (\ref{model:separable}). Then the following two statements are equivalent.
\begin{enumerate}[label=(\alph*)]
    \item The model satisfies \ref{assm:injective_general} (Injective) and \ref{assm:sym_general} (Symmetry).
    \item $V(x|\{x,y\}) = V(y|\{y,x\})$ if and only if $x_{1}=-x_{2}$.
\end{enumerate}
The result holds even when the alternatives have additional attributes with constants, for example, $x=(x_{1},x_{2},c_{3},\dots,c_{N})$ and $y=(0,0,c_{3},\dots,c_{N})$ in which $c_{3},\dots,c_{N} \in \mathbb{R}$.
\end{theorem}

\begin{proof}[Proof of Theorem~\ref{thm:pMPL12}]
[(a)$\implies$(b)]
By definition, $V(x|\{x,y\})=V(y|\{y,x\})$ is equivalent to $v(x_{1},0) + v(x_{2},0) = 0$. After rearranging it, this can be rewritten as $v^{0}(x_{1}) = v^{0}(-x_{2})$ due to the symmetry assumption of \ref{assm:sym_general}. By the injective assumption of \ref{assm:injective_general}, we conclude that $x_{1} = -x_{2}$. Next, suppose that $x_{1} = -x_{2}$. Then we have $v^{0}(x_{1}) = v^{0}(-x_{2})$, which implies that $V(x|\{x,y\})=V(y|\{y,x\})$.

[(b)$\implies$(a)]
First, we show that \ref{assm:sym_general} holds. Let $t\in\mathbb{R}$ be given. Consider $x=(t,-t)$ and $y=(0,0)$. By (b), it follows that $v(t,0)+v(-t,0) = V(x|\{x,y\}) = V(y|\{y,x\}) = 0$. This implies that $v^{0}(t) = -v^{0}(-t)$. Next, we show that \ref{assm:injective_general} holds. 
%Claim that, if $t>0$, then $v^{0}(t)>0$. To prove the claim, suppose to the contrary that $v^{0}(t_{0})=0$ for some $t_{0}>0$. Consider $x=(t_{0},0)$ and $y=(0,0)$. Then $V(x|\{x,y\}) = v^{0}(t_{0}) = 0$. Since we also have $V(y|\{y,x\})=0$, it follows that $t_{0}=0$ by (b). This is a contradiction. Now, 
Let $t>s>0$ be given. Consider $x=(t,-s)$ and $y=(0,0)$. By (b), we have $v^{0}(t)+v^{0}(-s) \neq 0$. Since we know that \ref{assm:sym_general} holds, we conclude that $v^{0}(t) \neq v^{0}(s)$.

The same logic can be applied to prove the case when the alternatives have additional attributes with constants as the model is additively separable. 
%Specifically, whenever additional attribute with constant $c\in\mathbb{R}$ is added to both alternatives, any separable attribute model adds an equal value of $v(c,c)$ to both evaluations of alternatives.
\end{proof}

A set of two-attribute alternatives above generalizes the p-MPL decision problem in the first scenario, i.e., when subjects ignore the initial endowment. A set of three-attribute alternatives with a constant in an additional attribute above generalizes the p-MPL decision problem in the second scenario, i.e., when subjects consider the initial endowment but separate it from price. Thus, regarding the quality elicitation problem, the result indicates $q=p^{*}$ in those cases. Corollary~\ref{cor:pMPL12_general} records the implication of Theorem~\ref{thm:pMPL12} that the injective and the symmetry assumptions guarantee that p-MPL accurately elicits quality when subjects ignore the initial endowment or consider it but separate it from price.

\begin{cor}\label{cor:pMPL12_general}
Suppose that a subject chooses according to a separable attribute model of (\ref{model:separable}) satisfying \ref{assm:injective_general} (Injective) and \ref{assm:sym_general} (Symmetry). 
\begin{enumerate}
    \item If the subject ignores the initial endowment so considers quality and price attributes, then p-MPL accurately elicits quality. Formally,
    \begin{align*}
        V((q,-p^{*})|\{(q,-p^{*}),(0,0)\}) &= V((0,0)|\{(0,0),(q,-p^{*})\})\\
        \text{ if and only if } q&=p^{*}.
    \end{align*}
    \item If the subject treats the initial endowment separately from price and quality so considers quality, price, and endowment attributes, then p-MPL accurately elicits quality. Formally,
    \begin{align*}
        V((q,-p^{*},E)|\{(q,-p^{*},E),(0,0,E)\}) &= V((0,0,E)|\{(0,0,E),(q,-p^{*},E)\})\\
        \text{ if and only if } q&=p^{*}.
    \end{align*}
\end{enumerate}
\end{cor}

Lastly, we examine the third scenario case when subjects combine the initial endowment and price in a single attribute. Theorem~\ref{thm:pMPL3} generalizes Proposition~\ref{prop:pMPL3} which shows that given two-attribute alternatives where one of them has a zero attribute value, a separable attribute model links the sum of attribute values of each alternative equivalently under the injective and the linearity assumptions.

\begin{theorem}\label{thm:pMPL3}
Let two-attribute alternatives $x=(x_{1},x_{2})$ and $y=(0,y_{2})$ with $x_{1},x_{2},y_{2}>0$ and $y_{2}>x_{2}$ be given. Suppose that a decision maker chooses according to a separable attribute model of (\ref{model:separable}). Then the following two statements are equivalent.
\begin{enumerate}[label=(\alph*)]
    \item The model satisfies \ref{assm:injective_general} (Injective) and \ref{assm:linear_general} (Linearity).
    \item $V(x|\{x,y\}) = V(y|\{y,x\})$ if and only if $x_{1}+x_{2}=y_{2}$.
\end{enumerate}
The result holds even when the alternatives have additional attributes with constants, for example, $x=(x_{1},x_{2},c_{3},\dots,c_{N})$ and $y=(0,y_{2},c_{3},\dots,c_{N})$ in which $c_{3},\dots,c_{N} \in \mathbb{R}$.
\end{theorem}

\begin{proof}[Proof of Theorem~\ref{thm:pMPL3}]
[(a)$\implies$(b)]
By definition, $V(x|\{x,y\}) = V(y|\{y,x\})$ is equivalent to $v(x_{1},0) + v(x_{2},y_{2}) = v(y_{2},x_{2})$. After rearranging it, this can be rewritten as $v^{0}(x_{1}) = v^{0}(y_{2}-x_{2})$ due to the linearity assumption of \ref{assm:linear_general}. By the injective assumption of \ref{assm:injective_general}, we conclude that $x_{1} + x_{2} = y_{2}$. Next, suppose that $x_{1} + x_{2} = y_{2}$. Then we have $v^{0}(x_{1}) = v^{0}(y_{2}-x_{2})$, which implies that $V(x|\{x,y\}) = V(y|\{y,x\})$.

[(b)$\implies$(a)]
We first show that \ref{assm:linear_general} holds. Let $t>s>0$ be given. Consider $x=(t-s,s)$ and $y=(0,t)$. By (b), we have $v(t-s,0) + v(s,t) = V(x|\{x,y\}) = V(y|\{y,x\}) = v(t,s)$. By rearranging it, we obtain $v^{0}(t-s) = v(t,s) - v(s,t)$. Next, we show that \ref{assm:injective_general} holds. Let $t>s>0$ be given. Consider $x=(t,\frac{1}{2}s)$ and $y=(0,\frac{3}{2}s)$. By (b), it follows that $v^{0}(t) + v(\frac{1}{2}s,\frac{3}{2}s) =V(x|\{x,y\}) \neq V(y|\{y,x\}) = v(\frac{3}{2}s,\frac{1}{2}s)$. Since we know that \ref{assm:linear_general} holds, we conclude that $v^{0}(t) \neq v(\frac{3}{2}s,\frac{1}{2}s) -  v(\frac{1}{2}s,\frac{3}{2}s) = v^{0}(s)$.

The same logic can be applied to prove the case when the alternatives have additional attributes with constants as the model is additively separable.
\end{proof}

A set of two-attribute alternatives above generalizes the p-MPL decision problem in the third scenario, i.e., when subjects combine the initial endowment and price in a single attribute. Corollary~\ref{cor:pMPL3_general} records the implication of Theorem~\ref{thm:pMPL3} that the injective and the linearity assumptions guarantee that p-MPL accurately elicits quality when subjects combine the initial endowment with price.

\begin{cor}\label{cor:pMPL3_general}
Suppose that a subject chooses according to a separable attribute model of (\ref{model:separable}) satisfying \ref{assm:injective_general} (Injective) and \ref{assm:linear_general} (Linearity). If the subject combines the initial endowment and price in a single attribute so considers quality and earning attributes, then p-MPL accurately elicits quality. Formally, 
\begin{align*}
    V((q,E-p^{*})|\{(q,E-p^{*}),(0,E)\} &= V((0,E)|\{(0,E),(q,E-p^{*})\})\\
    \text{ if and only if } q&=p^{*}.
\end{align*}
\end{cor}

\newpage
\section{More Examples}\label{appendix:example}
In this section, we apply the quality elicitation results to (weighted) separable attribute models other than the range normalization models studied in Examples~\ref{ex:RN_kinked} and \ref{ex:RN_power}. Example~\ref{ex:PN} applies the results to the pairwise normalization models studied by \cite{landry2021pairwise} and \cite{daviet2023test}. Example~\ref{ex:NCC} applies the results to the contextual concavity models studied by \cite{kivetz2004alternative}.

\begin{example}[Pairwise normalization models]\label{ex:PN}
A basic pairwise normalization (PN) model proposed by \cite{landry2021pairwise} evaluates alternative $x$ in menu $\{x,y\}$ by
\begin{equation}\label{model:PN}
    V(x|\{x,y\}) = \sum_{n=1}^{N} \frac{x_{n}}{|x_{n}| + |y_{n}|},
\end{equation}
where the evaluation is zero whenever two attribute values in the same attribute are zeros. As the model has a utility function of $u(t)=t$ and a weight function of $w(t,s) = \frac{1}{|t|+|s|}$, it is nested in a class of weighted separable attribute models of (\ref{model:weighted}). Since $u^{0}(t)=1$ for $t>0$, the injective assumption is violated. Hence, we can not use both m-MPL and p-MPL to accurately elicit quality for model (\ref{model:PN}). 

\cite{landry2021pairwise} also propose two types of generalized pairwise normalization models. One generalized pairwise normalization (GPN) model evaluates $x$ in $\{x,y\}$ by
\begin{equation}\label{model:GPN}
    V(x|\{x,y\}) = \sum_{n=1}^{N} \frac{x_{n}}{\sigma+|x_{n}| + |y_{n}|}
\end{equation}
where $\sigma>0$. As the model has a utility function of $u(t)=t$ and a weight function of $w(t,s) = \frac{1}{\sigma+|t|+|s|}$, it is nested in a class of weighted separable attribute models of (\ref{model:weighted}). The associated $u^{0}(t)=\frac{t}{\sigma+|t|}$ for all $t\in\mathbb{R}$ satisfies the injective and the symmetry assumptions. Thus, Corollaries~\ref{cor:mMPL} and \ref{cor:pMPL12} are applicable. However, we cannot apply Corollary~\ref{cor:pMPL3} as the linearity assumption is violated. 
%Specifically, for any $t_{1}$ and $t_{2}$, we have $v(t_{1}-t_{2}) = \frac{t_{1}-t_{2}}{\sigma+|t_{1}-t_{2}|}$ and $u(t_{1})w(t_{1},t_{2}) - u(t_{2})w(t_{1},t_{2}) = \frac{t_{1}-t_{2}}{\sigma+|t_{1}|+|t_{2}|}$. 
Furthermore, we find that $V((q,E-p^{*})|\cdot) = V((0,E)|\cdot)$ where the dots refer to corresponding menus implies $q=\frac{\sigma p^{*}}{\sigma+2E-2p^{*}}<p^{*}$ for all $p^{*}\in(0,E)$ and $\sigma>0$. This indicates that given model (\ref{model:GPN}), the switch point elicited by p-MPL is always greater than the model-implied quality in the third scenario case. For model (\ref{model:GPN}), we conclude that m-MPL accurately elicits quality, and p-MPL accurately elicits quality only when subjects ignore the initial endowment or separate it from price.

Another type of generalized pairwise normalization model, which is also empirically studied in \cite{daviet2023test}, is a pairwise normalization model with a power utility function (GPN w/ power utility). This model evaluates $x$ in $\{x,y\}$ by
\begin{equation}\label{model:GPNpower}
    V(x|\{x,y\}) =
    \sum_{n=1}^{N} \frac{u(x_{n})}{\sigma^{\alpha}+|x_{n}|^{\alpha} + |y_{n}|^{\alpha}}\\
\end{equation}
where $\sigma>0$, $\alpha>0$, and $u:\mathbb{R}\rightarrow\mathbb{R}$ is a power utility function as defined in Example~\ref{ex:RN_power}. The model has a weight function such as $w(t,s) = \frac{1}{\sigma^{\alpha} + |t|^{\alpha} + |s|^{\alpha}}$. Hence, it is nested in a class of weighted separable attribute models of (\ref{model:weighted}). The associated $u^{0}(t)=\frac{t^{\alpha}}{\sigma^{\alpha} + |t|^{\alpha}}$ for all $t\in\mathbb{R}$ satisfies the injective and the symmetry assumptions so Corollaries~\ref{cor:mMPL} and \ref{cor:pMPL12} are applicable. As the linearity assumption is violated,
%since $v(t_{1}-t_{2}) = \frac{(t_{1}-t_{2})^{\alpha}}{\sigma^{\alpha}+|t_{1}-t_{2}|^{\alpha}}$ and $u(t_{1})w(t_{1},t_{2}) - u(t_{2})w(t_{2},t_{1}) = \frac{t_{1}^{\alpha} - t_{2}^{\alpha}}{\sigma^{\alpha} + |t_{1}|^{\alpha} + |t_{2}|^{\alpha}}$. 
we cannot apply Corollary~\ref{cor:pMPL3} in this case. For model (\ref{model:GPNpower}), we conclude that m-MPL accurately elicits quality, and p-MPL accurately elicits quality only when subjects ignore the initial endowment or separate it from price.
\end{example}

The following example applies our main results to the contextual concavity model and normalized contextual concavity model \citep{kivetz2004alternative}. 

\begin{example}[Contextual Concavity Models]\label{ex:NCC}
Define a menu-dependent utility function by 
\begin{equation*}
    u_{CC}(t,s) = 
    \begin{cases}
    (t - \min\{t,s\})^{\theta} \quad &\text{for } t_,s\geq0\\
    -(|t| - \max\{t,s\})^{\theta}  \quad &\text{for } t,s\leq0
    \end{cases},
\end{equation*}
where $\theta\in(0,1)$. A contextual concavity (CC) model evaluates $x$ in $\{x,y\}$ by 
\begin{equation}\label{model:CC}
\begin{split}
    V(x|\{x,y\}) 
    &= \sum_{n=1}^{N} u_{CC}(x_{n},y_{n})
\end{split}.
\end{equation}
Consider that $v(t,s)=u_{CC}(t,s)$ for all $t,s\in\mathbb{R}$. Then we have $v(0,t)=0$, $v^{0}(t)=t^{\theta}$ for $t\geq 0$, and $v^{0}(t)=-(-t)^{\theta}$ for $t<0$. Thus, model (\ref{model:CC}) is nested in a class of separable attribute models of (\ref{model:separable}). Since $v^{0}(t)=t^{\theta}$ is strictly increasing, the injective assumption is satisfied. Since $v^{0}(t)=t^{\theta}=-v^{0}(-t)$ for all $t>0$, the symmetry assumption is satisfied. For any $t>s>0$, we have $v^{0}(t-s)=(t-s)^{\theta} = v(t,s) = v(t,s)-v(s,t)$ because $v(s,t)=(s-s)^{\theta}=0$. Thus, the linearity assumption is satisfied. Therefore, Corollaries~\ref{cor:mMPL_general}, \ref{cor:pMPL12_general}, and \ref{cor:pMPL3_general} are all applicable. For model (\ref{model:CC}), we conclude that both m-MPL and p-MPL accurately elicit quality.

A normalized contextual concavity (NCC) model evaluates $x$ in $\{x,y\}$ by
\begin{equation}\label{model:NCC}
\begin{split}
    V(x|\{x,y\}) 
    &= \sum_{n=1}^{N} u_{CC}(x_{n},y_{n})|x_{n}-y_{n}|^{1-\theta}
\end{split}.
\end{equation}
Here, model (\ref{model:NCC}) has a weight function such as $w(t,s)=|t-s|^{1-\theta}$. As we know the menu-dependent utility function satisfies $u_{CC}(0,t)=0$, is weakly increasing in the first argument, and is bounded, the model is nested in a class of separable attribute models of (\ref{model:separable}). Note that $v^{0}(t)=t$. Hence, the model satisfies all the assumptions, which implies that Corollaries~\ref{cor:mMPL_general}, \ref{cor:pMPL12_general}, and \ref{cor:pMPL3_general} are all applicable. For model (\ref{model:NCC}), we conclude that both m-MPL and p-MPL accurately elicit quality.
\end{example}

\begin{table}
\begin{center}
\caption{Quality elicitation results for models (\ref{model:PN}), (\ref{model:GPN}), (\ref{model:GPNpower}), (\ref{model:CC}), and (\ref{model:NCC})}\label{tab:PNNCC}
\begin{tabular}{lccc}
\toprule
MPL type & m-MPL & p-MPL & p-MPL \\
\text{[Scenario]} & $\cdot$ & \text{[Ignore or Separate]} & \text{[Combine]} \\
\midrule
PN & $\times$ & $\times$ & $\times$ \\
GPN & $\checkmark$ & $\checkmark$ & $\times (\uparrow)$ \\
GPN w/ power utility  & $\checkmark$ & $\checkmark$ & $\times$ ($\downarrow$ or $\uparrow$) \\ 
CC & $\checkmark$ & $\checkmark$ & $\checkmark$ \\
NCC & $\checkmark$ & $\checkmark$ & $\checkmark$ \\
\bottomrule
\end{tabular}
\end{center}
\vspace{3mm}
\footnotesize{\textit{Notes:} Scenario denotes how subjects treat the initial endowment during the decision-making process. The up arrow ($\uparrow$) indicates that the switch point is greater than the model-implied quality. The down arrow ($\downarrow$) indicates that the switch point is lower than the model-implied quality.}
\end{table}

Table~\ref{tab:PNNCC} summarizes the results from Examples~\ref{ex:PN} and \ref{ex:NCC}. First, both m-MPL and p-MPL cannot accurately elicit quality for model (\ref{model:PN}). This implies that we need to consider other types of elicitation methods to accurately elicit subjective quality for model (\ref{model:PN}). Second, for models (\ref{model:GPN}) and (\ref{model:GPNpower}), we need additional assumptions that subjects ignore the initial endowment or consider it but separate it from price to use p-MPL for accurate quality elicitation. In contrast, m-MPL accurately elicits quality without such additional assumptions. Lastly, for models (\ref{model:CC}) and (\ref{model:NCC}), both m-MPL and p-MPL can accurately elicit quality. Aligned with the discussion in Section~\ref{sec:discuss_somerville}, the examples here confirm the advantage of using m-MPL when testing multiple multi-attribute choice models following the experimental design of \cite{somerville2022range}.

\newpage

\section{Sign Test}\label{appendix:sign}
\subsection{Median test}
In this subsection, we study how to conduct the sign test to check whether the difference of the individual switch point, $m_{i}-p_{i}$, from a median subject is different from zero. 

The null hypothesis is $H_{0}: \Pr [m_{i}^{*}>p_{i}^{*}] = \Pr [p_{i}^{*}>m_{i}^{*}] = 0.5$, where the asterisk indicates the true switch point. Let $I$ denote the total number of subjects and $I_{m}$ denote the number of subjects who reported $m_{i}>p_{i}$ in the individual switch point data. Using the binomial distribution with a $50\%$ success rate, the likelihood of observing at least $I_{m}$ among $I$ subjects is 
\begin{equation}\label{eq:sign}
    G(I,I_{m})
    = \left( \frac{1}{2} \right)^{I} + {I \choose I-1} \left( \frac{1}{2} \right)^{I-1} \left( \frac{1}{2} \right)^{1} + \cdots + {I \choose I_{m}} \left( \frac{1}{2} \right)^{I_{m}} \left( \frac{1}{2} \right)^{I-I_{m}},
\end{equation}
in which it is the sum of $I-I_{m}+1$ number of terms. 

Let $I_{p}$ denote the number of subjects who reported $p_{i}>m_{i}$ in the individual switch point data. Then the $p$-value for a two-sided sign test under the null is $2\cdot G(I,\max\{I_{m},I_{p}\})$. In our dataset, we have $I=85$, $I_{m}=62$, and $I_{p}=23$. Thus, the $p$-value is $2\cdot G(85,62) < 0.0001$.

\subsection{Threshold computation}
In this subsection, we study how to compute the thresholds when classifying types using the product-specific switch point data. Recall that if a subject's m-score or p-score is greater than or equal to a threshold, then it means that the subject statistically behaves differently from a random subject who randomly makes decisions.

For each subject $i$, let $K_{i}$ be the number of positive-value products with distinct product-specific switch points in m-block and p-block, i.e., the number of product $j$'s such that $m_{i,j}\neq p_{i,j}$. In other words, $K_{i}$ indicates the number of products that are not nonpositive-value products and are not equal-value products. Recall that subject $i$'s \textit{absolute m-score} is the number of products in which $m_{i,j}>p_{i,j}$ and \textit{absolute p-score} is the number of products in which $p_{i,j}>m_{i,j}$.

Our goal is to find a threshold score where we can statistically say that subject $i$ behaves differently from the random subject when the subject's score passes the threshold. To this end, we use equation (\ref{eq:sign}). Specifically, given $K_{i}$, the likelihood that the random subject's absolute score is higher than or equal to $k$ is $G(K_{i},k)$. Therefore, subject $i$'s threshold score at the $5\%$ significance level is the maximum $k$ satisfying $2\cdot G(K_{i},k)<0.05$. For example, if $K_{i}=30$, then the threshold is $21$. This means that an absolute m-score or p-score greater than or equal to 21 is unlikely to happen for the random subject. 

%Table~\ref{tab:threshold} lists the threshold scores for different $K_{i}$'s. 
Note that subjects with different $K_{i}$'s would likely have distinct threshold scores. In addition, a threshold score cannot be computed when $K_{i}$ is less than 6 because $2\cdot G(5,5)=0.0625$. In our dataset, there are two subjects whose $K_{i}$ is less than 6. In this case, we classify these subjects into neither of the types.

\iffalse
\begin{table}[]
\centering
\caption{Threshold scores}
\label{tab:threshold}
\begin{tabular}{ll}
\hline
$K_{i}$ & Threshold score \\ \hline
30      & 21              \\
29      & 20              \\
        & 19              \\
        & 18              \\
        & 17              \\
        & 16              \\
        & 15              \\
        & 14              \\
        & 13              \\
        & 12              \\
        & 11              \\
        & 10              \\
        & 9              \\
        & 8              \\
        & 7              \\
        & 6               \\ \hline
\end{tabular}
\end{table}
\fi

In the above computation, we exclude equal-value products such that $m_{i,j}=p_{i,j}>0.01$. It is straightforward to compute the threshold considering these products. Let $K_{0}$ be the number of equal-value products. Then the threshold at the 5\% significance level including equal-value products is the maximum $k$ satisfying $2\cdot G(K_{i}+K_{0},k)$.

\section{Supplementary to Experiment}\label{appendix:experiment}
\subsection{List of Snacks}
\begin{table}[H]
\caption{List of snacks}
\begin{center}
\begin{tabular}{ll}
\toprule
Product \# & Product Description \\ \midrule
1 & Bare Fugi \& Reds Apple Chips (3.4 oz)  \\ 
2 & Cheetos Crunchy (8.5 oz)  \\ 
3 & Cheezit Original (12.4 oz)  \\ 
4 & Chips Ahoy Real Chocolate Chip Cookies Original (13 oz)  \\
5 & Coke (12 fl oz)  \\

6 & Coke Zero Sugar (12 fl oz) \\ 
7 & Doritos Nacho Cheese (9.25 oz) \\ 
8 & Flipz Milk Chocolate Covered Pretzels (7.5 oz) \\ 
9 & Goldfish Cheddar (6.6 oz) \\ 
10 & Haribo Gummi Candy (8 oz) \\ \midrule

11 & Hershey's Milk Chocolate (1.55 oz) \\ 
12 & Ice Breaker Ice Cubes Peppermint Flavored (3.24 oz, 40 pieces) \\ 
13 & KIND Caramel Almond \& Sea Salt (1.4 oz) \\ 
14 & KIND Dark Chocolate Nuts \& Sea Salt (1.4 oz) \\ 
15 & Lay's Classic (8 oz) \\

16 & Lotus Biscoff Cookies (8.8 oz) \\ 
17 & Milano Cookies Double Dark Chocolate (7.5 oz) \\ 
18 & Milano Cookies Milk Chocolate (6 oz) \\ 
19 & M\&M's Milk Chocolate (3.14 oz) \\ 
20 & OREO Chocolate Sandwich Cookies (14.3 oz) \\ \midrule

21 & Pocky Chocolate Cream (2.47 oz) \\ 
22 & Pocky Strawberry Cream (2.47 oz) \\ 
23 & Pop-Tarts Frosted Cookies and Cr\`{e}me (13.5 oz, 8 Pop-Tarts) \\ 
24 & Pop-Tarts Frosted S'mores (13.5 oz, 8 Pop-Tarts) \\ 
25 & Pringles Original (5.2 oz) \\

26 & Pringles Sour Cream \& Onion (5.5 oz) \\ 
27 & Skittles Original (2.17 oz) \\ 
28 & Snickers (1.86 oz) \\ 
29 & Sprite (12 fl oz) \\ 
30 & Twix (1.79 oz) \\
\bottomrule
\end{tabular}
\end{center}
\vspace{3mm}
\footnotesize{\textit{Notes:} Product Description is the description the subjects actually read during the experiment. The order of the snacks is randomized across subjects.}
\end{table}

\subsection{Instructions}
\subsubsection{General Instruction}
Welcome to the experiment! Thank you for your participation. 
This page contains basic information about the experiment. \textbf{\textit{Please read it carefully.}} 

You will participate in two tasks throughout the experiment. For convenience, let's call them \textit{Task A} and \textit{Task B}. The order of the tasks will be randomized. So, some of you will start with Task A then Task B, while some will start with Task B then Task A. 
Specific instructions will be given to you on the screen before each task. Each task consists of several rounds, and you will make a choice in each round. Please remember that there are no ``right" or ``wrong" choices. Your preferences may be different from other participants, and as a result, your choices can be different. Please note that as in all experiments in Economics, the procedures are described fully and all payments are real.
If you have any questions regarding the instructions, please raise your hand. The researcher will come to you and answer your questions.

Payment Method: Once you complete all the tasks, you will move on to the \textit{payment stage} page. 
In the payment stage, the computer will randomly select \textit{one} round from the entire Tasks A and B. Your choice in that round will determine your payment. You will also get a show-up fee of \$5. Therefore, your final payment from this experiment is \textbf{``payment from the randomly selected round + \$5 (show-up fee)."}
After the payment is determined, a short questionnaire page will appear on your screen. You will receive the final payment after you complete the questionnaire.

Instructions for the first task will appear on the next page. Please click the ``Start" button below to continue.

\subsubsection{Instruction for m-Block}
\textbf{\textit{Please read this page carefully!}} I am going to ask you the following list of questions:

\begin{table}[H]
\begin{center}
\begin{tabular}{rcccc}
\toprule
\textbf{Question \#} &                        & \textbf{Option A}  &  & \textbf{Option B} \\ \midrule
$1$           & Would you rather have: & Cheeseburger & or & \$0.01   \\
$2$           & Would you rather have: & Cheeseburger & or & \$0.02   \\
$3$           & Would you rather have: & Cheeseburger & or & \$0.03   \\
   $\vdots$ & $\vdots$               & $\vdots$  & $\vdots$   & $\vdots$ \\
$999$         & Would you rather have: & Cheeseburger & or & \$9.99   \\
$1,000$       & Would you rather have: & Cheeseburger & or & \$10.00  \\ 
\bottomrule
\end{tabular}
\end{center}
\vspace{3mm}
%\caption{Example of the m-MPL method. We expect that a subject chooses Option A for some initial questions, switches to Option B at some question number, and chooses Option B for the remaining questions. A switch point is defined by a question number where the subject switches their decision from Option A to B. An initial endowment is not required when conducting an actual experiment. In the payment stage, if Option A is chosen, then the subject obtains the product. If Option B is chosen, then the subject obtains the corresponding dollars.}
\end{table}

In each question, you can either pick \textbf{Option A (Product)} or \textbf{Option B (Money)}. If this round where the product is Cheeseburger were randomly chosen for payment, I would randomly pick \textit{one} question and pay you the option you chose on that one question. For example, suppose that Question \#328 is randomly chosen. If you chose Option A, then you get \textbf{Cheeseburger}. If you chose Option B instead, then you get \textbf{\$3.28}. Each question is equally likely to be chosen for payment. Obviously, you have no incentive to lie on any question, because if that question gets chosen for payment then you would end up with the option you like less.

I assume you are going to choose Option A in at least the first few questions, but at some point switch to choosing Option B. So, to save time, just tell me at which \textbf{\textit{dollar value}} you would switch. I can then ``fill out" your answers to all 1,000 questions based on your switch point (choosing Option A for all questions before your switch point, and Option B for all questions at and after your switch point). I will still draw one question randomly for payment. Again, if you lie about your true switch point you might end up getting paid for an option that you like less.

Cheeseburger is an example. You will see several different products in the main choice task. A picture and a name of a product will be presented together. Short quizzes will be provided before the main choice task. Please click the ``Next" button below to continue.

\subsubsection{Instruction for p-Block}
\textbf{\textit{Please read this page carefully!}} In Task B, I will first give you the \textbf{endowment of \$10}. Note that this \$10 is related to Task B \textit{only}. Then I am going to ask you the following list of questions:
\begin{table}[H]
\begin{center}
\begin{tabular}{rcccc}
\toprule
\textbf{Q \#} &  & \textbf{Option X}  &  & \textbf{Option Y} \\ \midrule
$1$           & Would you rather have: & Cheeseburger by paying $\$0.01$ & or & Nothing   \\
$2$           & Would you rather have: & Cheeseburger by paying $\$0.02$ & or & Nothing   \\
$3$           & Would you rather have: & Cheeseburger by paying $\$0.03$ & or & Nothing   \\
   $\vdots$ & $\vdots$               & $\vdots$  & $\vdots$ & $\vdots$ \\
$999$         & Would you rather have: & Cheeseburger by paying $\$9.99$ & or & Nothing   \\
$1,000$       & Would you rather have: & Cheeseburger by paying $\$10.00$ & or & Nothing  \\ 
\bottomrule
\end{tabular}
\end{center}
\vspace{3mm}
%\caption{Example of the m-MPL method. We expect that a subject chooses Option A for some initial questions, switches to Option B at some question number, and chooses Option B for the remaining questions. A switch point is defined by a question number where the subject switches their decision from Option A to B. An initial endowment is not required when conducting an actual experiment. In the payment stage, if Option A is chosen, then the subject obtains the product. If Option B is chosen, then the subject obtains the corresponding dollars.}
\end{table}

In each question, you can either pick \textbf{Option X (Buy Product)} or \textbf{Option Y (Nothing)}. If this round where the product is Cheeseburger were randomly chosen for payment, I would randomly select \textit{one} question and pay you the option you chose on that one question. For example, suppose that Question \#328 is randomly chosen. If you chose Option X, then your payment is \textbf{Cheeseburger} \textit{and} \textbf{\$6.72 (=\$10$-$3.28)} as you buy the product. If you chose Option Y instead, then your payment is \textbf{\$10} as you keep the endowment. Each question is equally likely to be selected for payment. Obviously, you have no incentive to lie on any question, because if that question gets chosen for payment then you would end up with the option you like less.

I assume you are going to choose Option X in at least the first few questions, but at some point switch to choosing Option Y. So, to save time, just tell me at which \textbf{\textit{price}} you would switch. I can then ``fill out" your answers to all 1,000 questions based on your switch point (choosing Option X for all questions before your switch point, and Option Y for all questions at and after your switch point). I will still draw one question randomly for payment. Again, if you lie about your true switch point you might end up getting paid for an option that you like less.

Cheeseburger is an example. You will see several different products in the main choice task. A picture and a name of a product will be presented together. Short quizzes will be provided before the main choice task. Please click the ``Next" button below to continue.

\subsection{Screenshots}
\begin{sidewaysfigure}[!htbp]
\caption{Screenshots of comprehension question pages}\label{fig:quiz}
\begin{subfigure}[b]{0.49\textwidth}
    \includegraphics[scale=0.35]{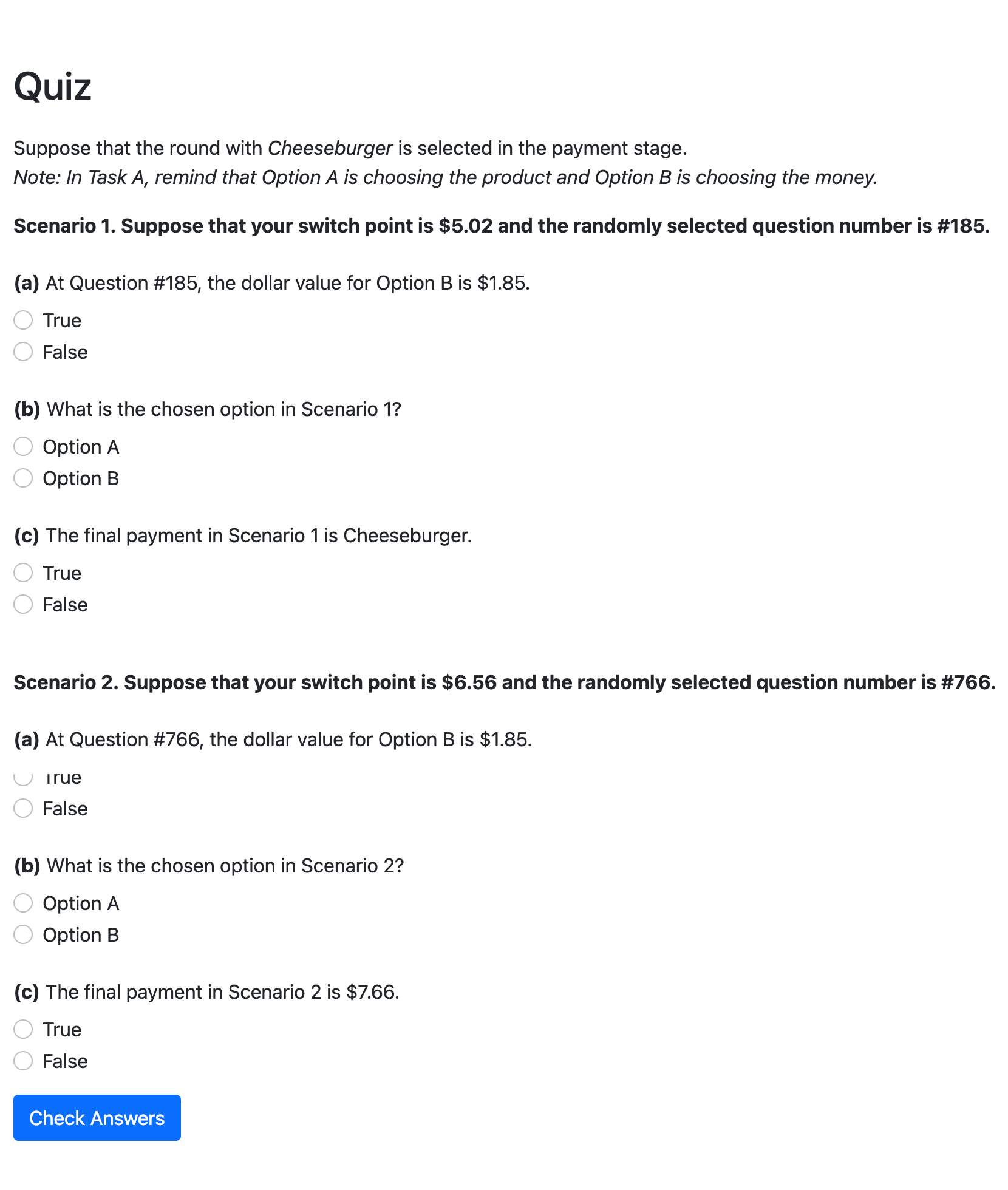}
    \caption{A comprehension question page in m-block}\label{fig:quiz_m}
\end{subfigure}
\begin{subfigure}[b]{0.49\textwidth}
    \includegraphics[scale=0.35]{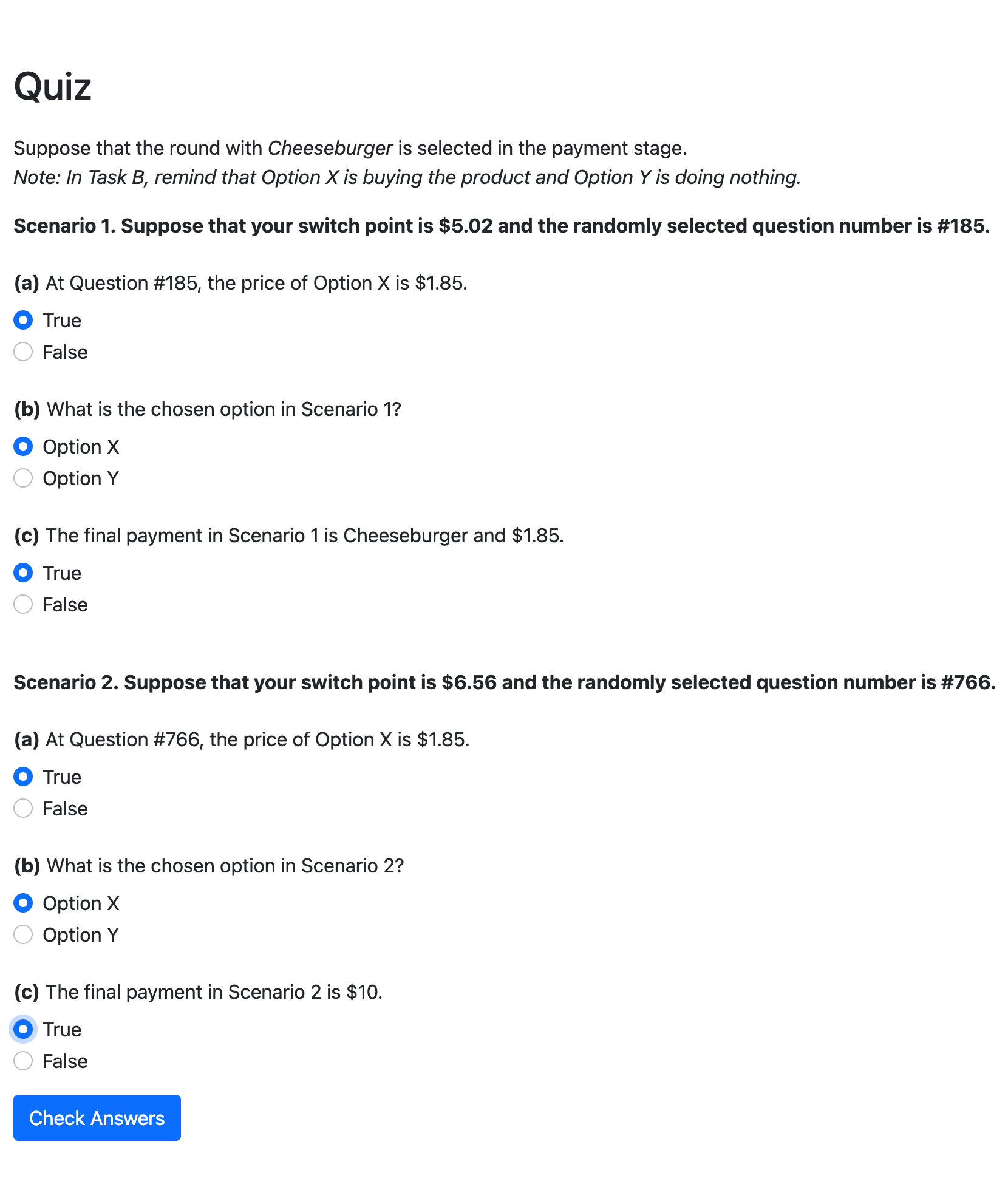}
    \caption{A comprehension question page in p-block}\label{fig:quiz_p}
\end{subfigure}
\end{sidewaysfigure}

\begin{sidewaysfigure}[!htbp]
\caption{Screenshots of comprehension question answer pages}\label{fig:quiz_answer}
\begin{subfigure}[b]{0.49\textwidth}
    \includegraphics[scale=0.3]{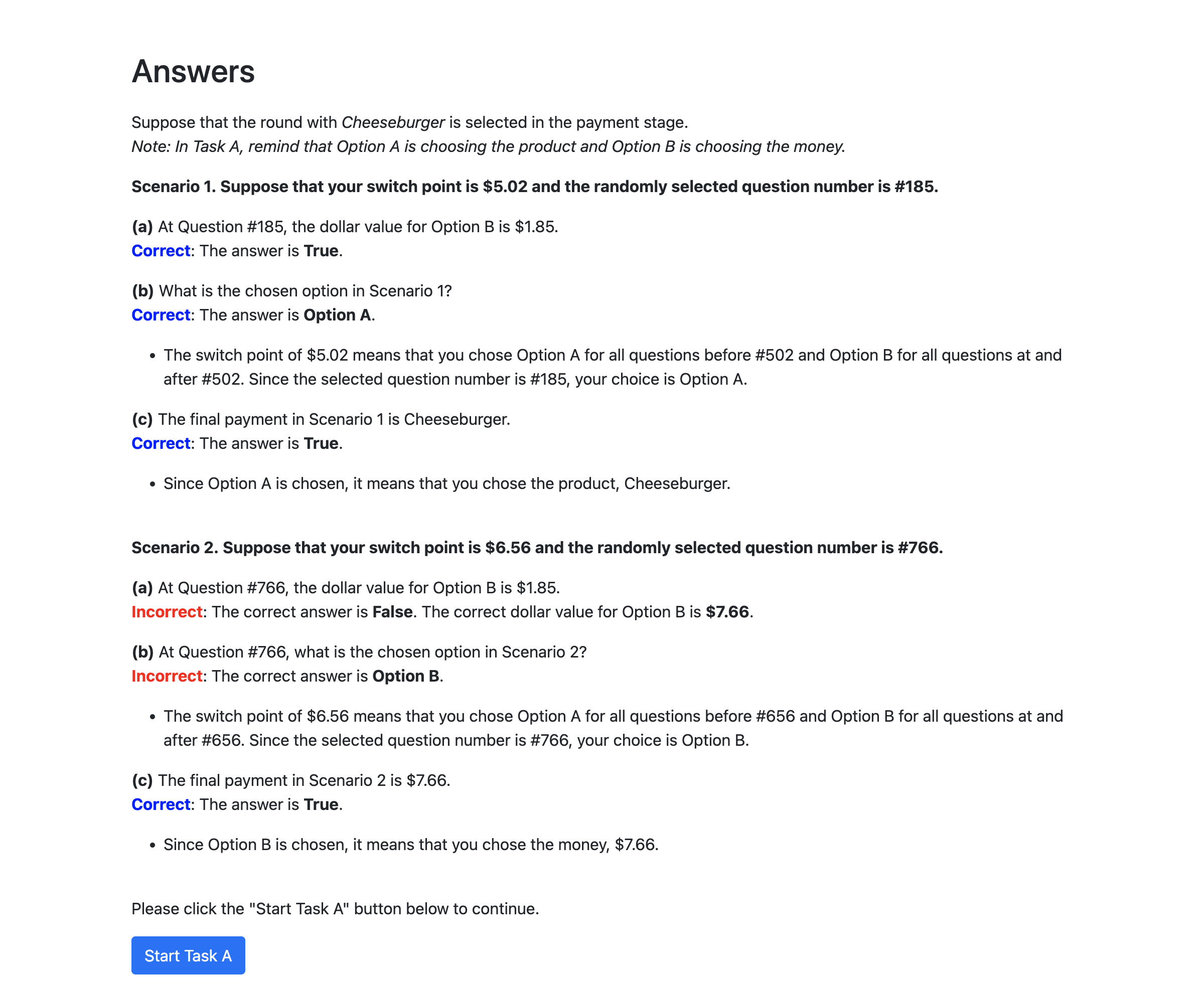}
    \caption{A comprehension question answer page in m-block}\label{fig:quiz_m_answer}
\end{subfigure}
\begin{subfigure}[b]{0.49\textwidth}
    \includegraphics[scale=0.3]{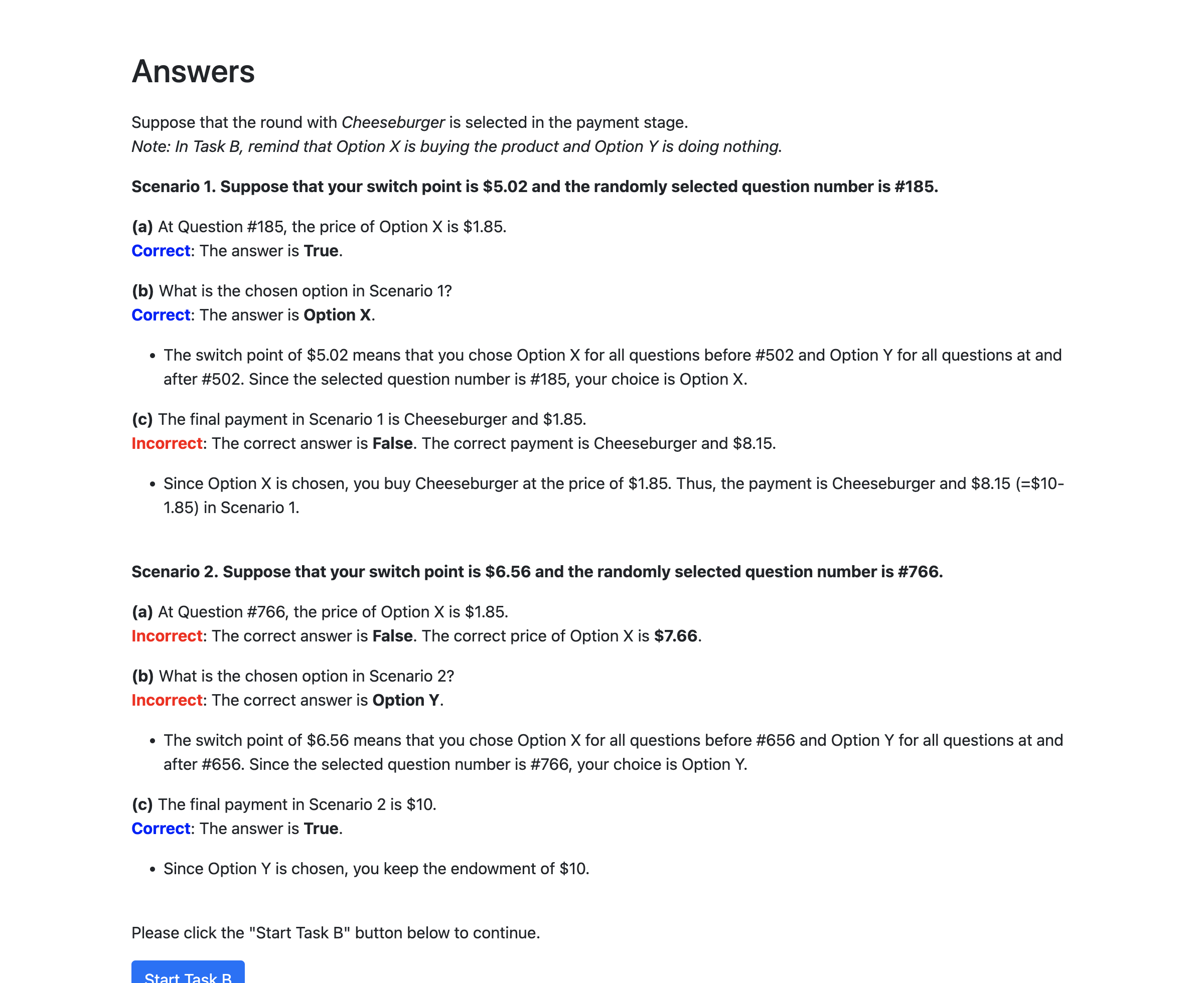}
    \caption{A comprehension question answer page in p-block}\label{fig:quiz_p_answer}
\end{subfigure}
\end{sidewaysfigure}

\newpage
\section{Supplementary to Data Analysis}\label{appendix:data}
\subsection{Demographic Information}
Table~\ref{tab:demo} summarizes the demographic information of subjects from mp-treatment and pm-treatment, respectively. Age, Gender, Race, and Major are self-reported by subjects at the end of the experiment. Demographic information of two subjects in mp-treatment is not collected by mistake. Comprehension questions are presented after subjects read a set of specific instructions and before the main task. For the comprehension questions, Table~\ref{tab:demo} reports the average number of correct answers without missing data. 

\begin{table}[!htbp]
\caption{Subject Demographics}\label{tab:demo}
    \begin{center}
        \begin{tabular}{lcc}
        \toprule
         & mp-treatment & pm-treatment \\ \midrule
         Number of Subjects & 42 & 43 \\ 
         %\addlinespace
         \textit{Age} & 20.8 & 21 \\
         %Age & \begin{tabular}[c]{@{}c@{}}20.78 \\ (1.78)\end{tabular} & \begin{tabular}[c]{@{}c@{}}21.00\\ (2.00)\end{tabular} \\
         %\addlinespace
         \textit{Gender} & & \\
         \;\; Female & 60\% & 46.5\% \\
         \;\; Male & 40\% & 53.5\% \\ 
         %\addlinespace
         \textit{Race} & & \\ 
         \;\; African-American/Black & 5\% & 2.3\% \\
         \;\; American Indian or Alaskan Native & 0\% & 2.3\% \\
         \;\; Asian American/Asian & 25\% & 20.9\% \\
         \;\; Hispanic/Latino & 5\% & 9.3\% \\
         \;\; Multi-Racial & 0\% & 4.7\% \\
         \;\; White & 65\% & 60.5\% \\
         %\addlinespace
         \textit{Major} & & \\
         \;\; Business/Economics & 62.5\% & 62.8\% \\
         \;\; Non-Business/Economics & 37.5\% & 37.2\% \\
         %\addlinespace
         \textit{Comprehension Questions} & & \\
         \;\; m-block Questions Correct & 5.3 & 5.2 \\
         %\;\; m-block Questions Correct & \begin{tabular}[c]{@{}c@{}}5.29\\ (1.22)\end{tabular} & \begin{tabular}[c]{@{}c@{}}5.21\\ (1.30)\end{tabular} \\
         \;\; p-block Question Correct & 5.2 & 4.9 \\
         %\;\; p-block Question Correct & \begin{tabular}[c]{@{}c@{}}5.24\\ (0.96)\end{tabular} & \begin{tabular}[c]{@{}c@{}}4.91\\ (1.23)\end{tabular} \\
         \;\; Total Question Correct & 10.5 & 10.1 \\ 
         %Total Question Correct & \begin{tabular}[c]{@{}c@{}}10.52\\ (1.84)\end{tabular} & \begin{tabular}[c]{@{}c@{}}10.12\\ (1.75)\end{tabular} \\
         % & \begin{tabular}[c]{@{}c@{}}2.78\\ (1.15)\end{tabular} & \begin{tabular}[c]{@{}c@{}}1.81\\ (1.51)\end{tabular} \\  \hline
         %& \begin{tabular}[c]{@{}c@{}}2.95\\ (2.08)\end{tabular} & \begin{tabular}[c]{@{}c@{}}1.93\\ (1.08)\end{tabular} \\ \hline
         \bottomrule
        \end{tabular}
    \end{center}   
    \vspace{3mm}
    \footnotesize{\textit{Notes:} Demographic information of two subjects in mp-treatment is not collected by mistake during the experiment. Thus, \textit{Age}, \textit{Gender}, \textit{Race}, and \textit{Major} in mp-treatment are based on 40 subjects. For each treatment, six comprehension questions are given. Hence, subjects who submitted all correct answers got 12 points.}
\end{table}

\subsection{CDFs of the Nonpositive-value Products, Equal-value Products, and Normalized Scores}\label{appendix:data_cdf}
\begin{figure}[!htbp]
    \caption{CDFs}
    \begin{center}
    \begin{subfigure}[b]{0.48\textwidth}
        \includegraphics[scale=0.34]{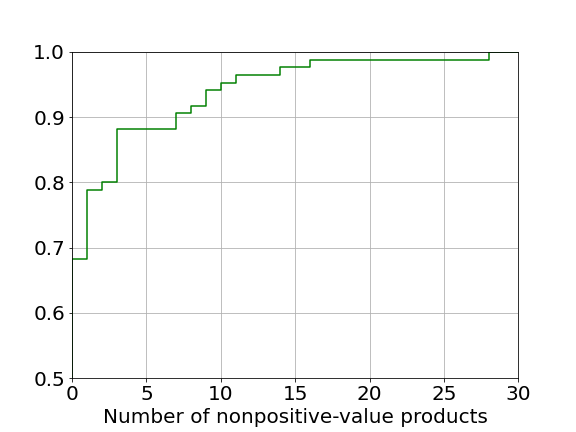}
        \caption{CDF of the number of nonpositive-value products}
        \label{fig:cdf_nonpositive}
    \end{subfigure}
    \begin{subfigure}[b]{0.48\textwidth}
        \includegraphics[scale=0.34]{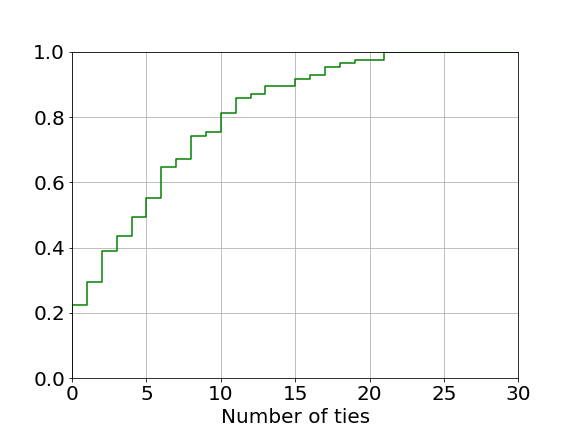}
        \caption{CDF of the number of equal-value products}
        \label{fig:cdf_tie}
    \end{subfigure}
    \end{center}
    
    \begin{subfigure}[b]{0.48\textwidth}
        \includegraphics[scale=0.35]{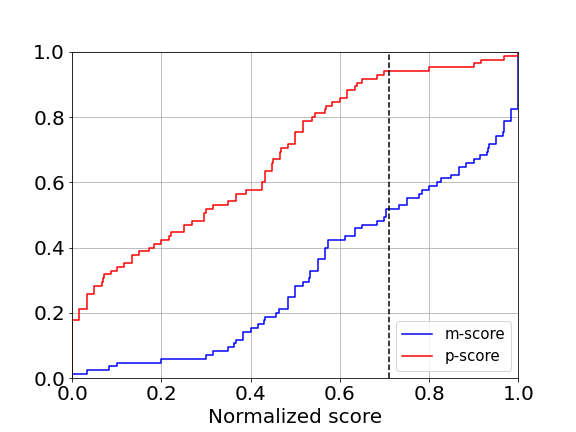}
        \caption{CDF of the scores when equal-value products are equally distributed}
        \label{fig:cdf_scores_equal}
    \end{subfigure}
    \begin{subfigure}[b]{0.48\textwidth}
        \includegraphics[scale=0.35]{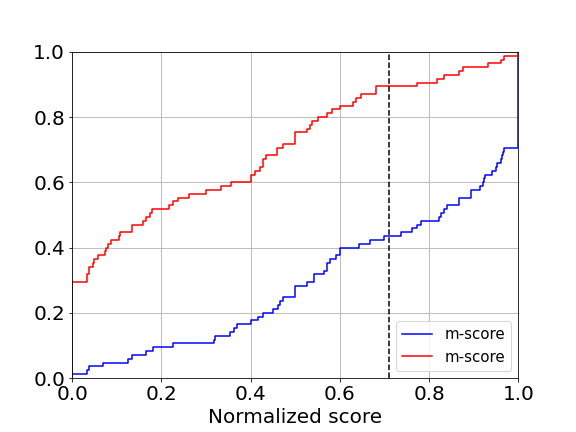}
        \caption{CDF of the scores when equal-value products are proportionally distributed}
        \label{fig:cdf_scores_prop}
    \end{subfigure} 
    \label{fig:cdfs}
    \footnotesize{\textit{Notes:} In Figure~\ref{fig:cdf_nonpositive}, a nonpositive-value product is a product with product-specific switch points equal to 0.01 in both blocks, i.e., $m_{i,j}=p_{i,j}=0.01$. In Figure~\ref{fig:cdf_tie}, an equal-value product is a positive-value product with equal product-specific switch points in both blocks, i.e., $m_{i,j}=p_{i,j}>0.01$. The vertical dotted line in Figures~\ref{fig:cdf_scores_equal} and \ref{fig:cdf_scores_prop} indicates the weighted average threshold of 0.71.}
\end{figure}

Figure~\ref{fig:cdf_nonpositive} plots a cumulative distribution function (CDF) of the number of nonpositive-value products, i.e., $m_{i,j}=p_{i,j}=0.01$. Roughly $70\%$ of subjects reported zero nonpositive-value products, meaning that they have 30 positive-value products in each block. Roughly 90\% of subjects reported nonpositive-value products smaller than 4, meaning that they have more than 26 positive-value products. One subject reported 28 nonpositive-value products.

Figure~\ref{fig:cdf_tie} plots a CDF of the number of the equal-value products, i.e., $m_{i,j}=p_{i,j}>0.01$. Note that equal-value products do not include nonpositive-value products since they are $m_{i,j}=p_{i,j}=0.01$. Around $20\%$ of subjects reported zero equal-value products, which means that, given a product, they always reported different product-specific switch points between the blocks. One subject reported 21 equal-value products.

Figure~\ref{fig:cdf_scores_equal} plots CDFs of the normalized scores when equal-value products are equally distributed to the scores. This approach provides a lower bound for each type as it would make subjects less likely to pass the threshold. Using this approach, 48\% of subjects are classified as m-high type whereas 7\% of subjects are classified as p-high type. 

Figure~\ref{fig:cdf_scores_prop} plots CDFs of the normalized scores when equal-value products are proportionally distributed to the scores based on the ratio of the scores. This approach provides an upper bound for each type as it would make subjects more likely to pass the threshold. Using this approach, 56\% of subjects are classified as m-high type whereas 11\% of subjects are classified as p-high type.

\newpage

\subsection{Robustness Check using the Entire Dataset}\label{appendix:data_entire}
For robustness check, we proceed with data analysis using the entire dataset that includes both positive-value and nonpositive-value products. In the entire dataset, there are 60 product-specific switch point choices for each subject.

First, we check whether the order effect of blocks exists using the individual switch point data. Note that the individual switch point using the entire dataset is computed by the sum of all product-specific switch points divided by 30, i.e., $\frac{1}{30} \sum_{j=1}^{30} m_{i,j}$ for m-block and $\frac{1}{30} \sum_{j=1}^{30} p_{i,j}$ for p-block. Table~\ref{tab:mean_entire} reports the means of individual switch points for each treatment and for each block. In m-block, the means are 2.68 in mp-treatment and 2.83 in pm-treatment, and the difference is insignificant ($p\text{-value}>0.73$, two-sided Wilcoxon rank-sum test). In p-block, the means are 1.75 in mp-treatment and 1.89 in pm-treatment, and the difference is insignificant ($p\text{-value}>0.26$, two-sided Wilcoxon rank-sum test). Thus, no order effect is detected, and we pool the data across treatments in the following analysis.

\begin{table}[!htbp]
\caption{Means of individuals switch points using the entire dataset}
\label{tab:mean_entire}
\begin{center}
\begin{tabular}{lcc}
\toprule
             & m-block  & p-block  \\
\midrule
mp-treatment & $2.68$   & $1.75$   \\
             & $(1.22)$ & $(1.52)$ \\
pm-treatment & $2.83$   & $1.89$   \\
             & $(2.15)$ & $(1.09)$ \\
\midrule
Pooled       & $2.76$   & $1.82$   \\
             & $(1.75)$ & $(1.32)$ \\
\bottomrule
\end{tabular}
\end{center}
\vspace{3mm}
    \footnotesize{\textit{Notes:} The table reports the means of individual switch points using the entire dataset including positive-value and nonpositive-value products. No order effect is detected. Standard deviations are in parentheses.}
\end{table}

Table~\ref{tab:mean_entire} reports that the mean individual switch points are 2.76 in m-block and 1.82 in p-block in the pooled dataset, and they are significantly different ($p$-values$<0.0001$, two-sided paired $t$-test and Wilcoxon signed-rank test). Consistent with the main findings, on average, we find that the subjective quality elicited by p-MPL is significantly lower than the quality elicited by m-MPL in the entire dataset. The entire dataset provides the same median test result as in Section~\ref{sec:experiment} because the number of subjects (62 out of 85) who reported a higher individual switch point when m-MPL is used compared to p-MPL does not change.

Table~\ref{tab:reg_entire} shows the regression results of the linear model using the entire dataset. Following the main analysis, column 1 reports the results without fixed effects, column 2 reports the results including the individual fixed effects only, and column 3 reports the results including the individual and product fixed effects.

\begin{table}
    \caption{Regression results using the entire dataset}\label{tab:reg_entire}
        \begin{center}
{
\def\sym#1{\ifmmode^{#1}\else\(^{#1}\)\fi}
%\begin{tabular}{l*{4}{D{.}{.}{-1}}}
\begin{tabular}{lcccc}
\toprule
                    & (1)              & (2)              & (3)               \\
\midrule
Block               & $-0.918\sym{***}$& $-0.918\sym{***}$& $-0.918\sym{***}$ \\
                    & $(0.214)$        & $(0.216)$        & $(0.216)$         \\
\addlinespace
Price               & $0.277\sym{***}$ & $0.277\sym{***}$ & -                 \\
                    & $(0.027)$        & $(0.028)$        & -                 \\
\addlinespace
Block$\times$Price  & $-0.008$         & $-0.008$         & $-0.008$          \\
                    & $(0.022)$        & $(0.022)$        & $(0.022)$          \\
\addlinespace
Constant            & $2.036\sym{***}$ & $1.159\sym{***}$ & $2.116\sym{***}$  \\
                    & $(0.201)$        & $(0.139)$        & $(0.194)$         \\
\midrule
Individual fixed effects & No          & Yes              & Yes         \\
Product fixed effects    & No          & No               & Yes         \\
\midrule
Observations        & $5,100$          & $5,100$          & $5,100$          \\
Number of subjects  & $85$             & $85$             & $85$             \\
\bottomrule
\end{tabular}
}

\end{center}
\vspace{3mm}
\footnotesize{\textit{Notes:} The dependent variable is $Switch_{i,j,b}$ denoting subject $i$'s switch point for product $j$ in $b$-block. $Block_{b}$ denotes a block indicator where $Block_{m}=0$ and $Block_{p}=1$. $Price_{j}$ denotes the market price of product $j$. A coefficient of $Price$ is omitted when the product fixed effects are included as the market price is invariant for each product.
Clustered standard errors at the individual level are in parentheses. \\ * $p<0.10$, ** $p<0.05$, *** $p<0.01$.}
\end{table}

First, all the coefficients of $Block$ are negative and significant, which implies that the quality elicited by p-MPL is lower than the quality elicited by m-MPL. Second, while the coefficients of $Price$ in columns 1 and 2 are positive and significant, it is less than one, indicating that the quality measure is not driven entirely by price. Third, the coefficients of $Block \times Price$ in columns 1, 2, and 3 are close to zero and insignificant. This implies that while subjects are sensitive to prices when making switch point decisions, price sensitivity does not differ depending on the MPL used in the experiment. These are all consistent with the main findings.

\newpage

\bibliographystyle{chicago}
\bibliography{ref}

\end{document}